\documentclass[11pt,a4paper]{article}
\pdfoutput=1

\usepackage{jheppub}
\usepackage[T1]{fontenc}
\usepackage{lmodern}
\usepackage{microtype}
\usepackage{mathtools}
\usepackage{bm}
\usepackage{booktabs}
\usepackage{array}
\usepackage{enumitem}
\usepackage{xcolor}
\usepackage{slashed}
\usepackage{tikz}
\usetikzlibrary{arrows.meta,calc,decorations.pathreplacing,fit,positioning,shapes.geometric}

\newcommand{\Npg}{N_{\mathrm{pg}}}

\makeatletter
\newcommand{\equationaliases}[1]{%
  \begingroup\protected@edef\@currentlabel{\theequation}#1\endgroup}
\makeatother
\newcommand{\Nc}{N_c}
\newcommand{\GN}{G_N}
\newcommand{\Hc}{\mathcal H_C}

\newcommand{\VC}{V_C}
\newcommand{\GC}{G_C}
\newcommand{\WC}{W_C}

\newcommand{\id}{\operatorname{id}}
\newcommand{\Tr}{\operatorname{Tr}}
\newcommand{\rank}{\operatorname{rank}}

\newcommand{\Prob}{\mathbb P}
\newcommand{\cN}{\mathcal N}
\newcommand{\cR}{\mathcal R}
\newcommand{\cV}{\mathcal V}

\definecolor{bookletblue}{RGB}{51,102,170}
\definecolor{bookletorange}{RGB}{221,119,51}
\definecolor{bookletgreen}{RGB}{68,153,112}
\definecolor{bookletpurple}{RGB}{136,85,170}
\definecolor{bookletgray}{RGB}{95,99,104}
\renewcommand{\afterTocRuleSpace}{\clearpage}
\usepackage{eso-pic}
\newcommand{\reportnum}[2]{
  \AddToShipoutPictureBG*{%
    \AtPageUpperLeft{%
      \hspace{0.75\paperwidth}%
      \raisebox{#1\baselineskip}{%
        \makebox[0pt][l]{\textnormal{#2}}
  }}}%
}

\reportnum{-6}{USTC-ICTS/PCFT-26-67}

\title{A baby universe from a large family: booklet cosmology states and quantum error correction}

\author[a,b]{Jingshu Dai}
\emailAdd{daijs@ihep.ac.cn}

\author[a,b]{, Binye Dong}
\emailAdd{dongby@ihep.ac.cn}

\author[a,c]{, and Cheng Peng}

\emailAdd{pengcheng@ucas.ac.cn}

\affiliation[a]{\small University of the Chinese Academy of Sciences, Beijing 100190, China}
\affiliation[b]{\small Institute of High Energy Physics, Chinese Academy of Sciences, Beijing 100049, China}
\affiliation[c]{\small Peng Huanwu Center for Fundamental Theory, Hefei, Anhui 230026, China}

\abstract{
We propose an extension of the cosmological-state construction of Antonini, Sasieta and Swingle (AS$^2$) to three or more holographic CFTs. Each CFT resides on the asymptotic boundary of an AdS page, and all pages are glued along a common interface by imposing the multiway junction conditions. The associated states are prepared by Euclidean evolution with a multilinear insertion. In appropriate heavy insertion limit the bulk geometry develop a closed universe in the center and we refer the state in this limit as a booklet cosmological state. Extending the effective Gaussian assumption used in AS$^2$, we model
the three-page insertion by a circular complex Gaussian random tensor, yielding a tripartite Haar state in flat energy windows. In this Gaussian model of the cosmology-to-boundary map, each arm is a network branch associated with one boundary CFT. In the simplest example of three pages with three equal output Hilbert-space dimension $b$, we show that a prescribed code of dimension $K$ is approximately recoverable from any two arms, with vanishing error and high probability as $K/b\to0$.
}

\begin{document}
\maketitle
\flushbottom
\section{Introduction and summary of results}
\label{sec:introduction}

AdS/CFT provides a framework for investigating the quantum
description of cosmological spacetimes
\cite{Maldacena:2004rf,Hertog:2005hu,Freivogel:2005qh,
Cooper:2018cmb,Antonini:2019qkt,VanRaamsdonk:2021qgv,
Antonini:2022ptt,Sahu:2023fbx,
Antonini:2022blk,Sahu:2024ccg}.
A central question in this direction is how quantum information in a closed universe is encoded in a boundary theory. An interesting recent proposal~\cite{ASS2023,Antonini:2025ioh} considered  Euclidean preparation with a heavy operator insertion that produces a partially entangled thermal state of two holographic CFTs. In an appropriate low-temperature semiclassical regime, the Lorentzian bulk contains a closed cosmological region, or baby universe, together with two asymptotically AdS regions. It is also showned that the encoding of such cosmological states in the joint boundary state can be tractably described using random tensor-network models.

In this work, we extend this construction to a fixed finite number $\Npg\geq 3$ of holographic CFTs, using Euclidean evolution with a multilinear insertion. The Euclidean bulk preparation has a ``booklet'' structure: asymptotically AdS pages meet along a common spherical shell, or spine, and are glued together by the multiway junction conditions~\cite{Shen:2024itl,ShenPengLi2024}.~\footnote{There have been a variaty of applications of the multi-way junction condition~\cite{Liu:2025khw,Guo:2025sbm,Jiang:2025iet,Chakraborty:2025jtj,Zhao:2025kru,Banerjee:2025zuw,Ogonowski:2025fgm,Guo:2026prm,Jiang:2026cui,Shen:2026coi,Liu:2026afo}
 and suggestions of potential applications of the booklet geometry to understand better different holographic dualities~\cite{Basteiro:2024crz,Yu:2025tid,Li:2025lpa,Cui:2025sgy,Afxonidis:2026sao}. } The low-temperature, large-mass saddle has a Lorentzian continuation containing a closed universe in the center and we refer the associated states ``booklet cosmological states''.

In this regime, we use replica calculations to estimate the entanglement entropy between boundary subsets and their complements, under appropriate assumptions on factorization,replica saddles and analytic continuation. We also define a cosmology-to-boundary map from local cosmological probe states that describes bulk matter excitations in the cosmology, to joint boundary CFT states. Extending the effective Gaussian assumption used in~\cite{ASS2023, Antonini:2025ioh}, we construct a random tensor-network model of this encoding and use it to study multipartite entanglement and approximate
quantum error correction. Our main findings are summarized below.

\paragraph{Booklet cosmological state and islands}

The booklet cosmological state is prepared by inserting a multilinear tensor $\cV$ in to the euclidean evolution with one
Euclidean preparation length per page. We adopt a thin-shell
interpretation and a multipage version of the generalized
eigenstate thermalization hypothesis (ETH) ansatz used in~\cite{ASS2023}. Section~\ref{sec:states-geometry}
describes the geometry and its gluing at the common shell.

For a boundary subset $A$, we extend the replica trick computation to the booklet geometry. In the large-mass factorization limit, under appropriate assumptions on replica-saddle dominance and a controlled continuation to one replica, we obtain the semiclassical estimate
\begin{equation}
 S_A^{\rm sc}\simeq
 \min\!\left\{S_{\rm bulk}(a_A),
 S_{\rm bulk}\bigl(a_A\sqcup c^{(\Npg)}\bigr)\right\}\ .
 \label{eq:intro-island-formula}
\end{equation}
Here $a_A=\bigsqcup_{i\in A}a_i$ is the union of the bulk exterior regions associated with the boundaries in $A$, $c^{(\Npg)}$ is the entire compact cosmological region, and $S_{\rm bulk}$ denotes the bulk matter entropy.

\paragraph{Random tensor model and quantum error correction}

At fixed Euclidean preparation, we define a linear map from local probe excitation states in the cosmology region to joint boundary states. In section~\ref{sec:tensor-network}, a multi-arm tensor network is defined, which provides a finite model of this map. The model consists of one fixed shell tensor characterizeing inputs of the shell and one arm per boundary CFT. With flat bonds and independent circular complex Gaussian coefficients, the normalized output for any fixed input is Haar distributed on the effective output space. In the simplest example of three equal arms in the large-dimension limit, these states have substantial pairwise mutual information, yet typically admit neither near-perfect one-copy EPR extraction by product local channels nor extraction of an exact GHZ factor by local unitaries~\cite{LiMoriYoshida2026}.

We also analyzed the quantum error correction properties of this map. For three arms with equal effective output dimension $b$, polar normalization of the Gaussian map on a prescribed $K$-dimensional code, $2\leq K\leq b^3$, gives a random isometric encoder~\cite{HSW2008,Klesse2007}. When $b\to\infty$ and $K/b\to0$, any two arms recover the logical state with vanishing worst-case error and with probability approaching one. At each finite $b$, however, exact correction of a prescribed one-arm erasure has probability zero. 
%
% ===== BEGIN sections/states_geometry =====
\section{Booklet cosmological states}
\label{sec:states-geometry}

The construction in~\cite[section~2]{ASS2023} produces two asymptotic AdS regions and a closed
cosmology.  We extend it by joining \(\Npg\)
AdS pages \(V_i\) along a common thin shell \(W\), called the spine.
The selected low-temperature geometry again prepares separate exterior
regions and a compact central region.  For \(\Npg\ge3\), the center is
a branched non-manifold rather than a smooth spacetime.  Its construction
therefore requires both a multiway junction condition and a choice of how
the pages are cut and identified.  We specify these below, together with
the assumptions on the bulk state needed for the replica calculation.

We work in \(D=d+1\) bulk dimensions, with \(d\ge2\) and \(\hbar=c=1\).
The AdS radius is \(\ell\), Newton's constant is \(\GN\), and
\(\Omega_{d-1}\) is the volume of the unit \((d-1)\)-sphere.
All pages have the same gravitational and matter couplings, but their
preparation lengths and quantum states may differ.  The page number
\(\Npg\) is finite and fixed throughout.  We reserve \(\Nc\) for the
large-rank parameter of the boundary CFTs.

\subsection{State preparation}
\label{sec:booklet-state}

For page \(i\), let \(\mathcal H_i\) be the CFT Hilbert space and \(H_i\)
its self-adjoint Hamiltonian.  We use energy eigenvectors
\(\lvert n_i;\epsilon_i\rangle_i\), with eigenvalues \(E_{i,n_i}\).
The sign \(\epsilon_i=+\) denotes a factor in \(\mathcal H_i\), whereas
\(\epsilon_i=-\) denotes its conjugate space \(\overline{\mathcal H}_i\).
Reversing a boundary cut's orientation replaces its Hilbert space by the
conjugate space.
Section~\ref{sec:booklet-saddle} relates this convention to the
Euclidean preparation contour.

Let \(\cV_{n_1\ldots n_{\Npg}}\) be the coefficients of one heavy
multilinear insertion.  Euclidean evolution over positive lengths
\(\widetilde\beta_i/2\) prepares the multipartite partially entangled
thermal state (multi-PETS)
\begin{align}
 \lvert\widehat\Psi_{\cV}\rangle
 &=\sum_{n_1,\ldots,n_{\Npg}}
 \cV_{n_1\ldots n_{\Npg}}
 \exp\!\left[-\frac12\sum_{i=1}^{\Npg}
       \widetilde\beta_i E_{i,n_i}\right]
 \bigotimes_{i=1}^{\Npg}\lvert n_i;\epsilon_i\rangle_i,
 \label{eq:sg-booklet-pets}\\
 \lvert\Psi_{\cV}\rangle&=Z_{\cV}^{-1/2}
 \lvert\widehat\Psi_{\cV}\rangle,\qquad
 Z_{\cV}=\sum_{\boldsymbol n}
 |\cV_{\boldsymbol n}|^2e^{-\sum_i\widetilde\beta_iE_{i,n_i}} .
 \label{eq:sg-booklet-norm}
\end{align}
Here \(\boldsymbol n=(n_1,\ldots,n_{\Npg})\), and the bra insertion has
coefficients \(\cV^*_{\boldsymbol n}\).
We assume \(0<Z_{\cV}<\infty\) at the chosen preparation lengths.
When truncating the energy sums, we use the same cutoff and adjoint
conventions in the state, its norm and all sewn copies.  For two pages,
\(\boldsymbol\epsilon=(-,+)\) and \(\cV_{mn}=\mathcal O_{mn}\), with
\(\mathcal O:\mathcal H_R\to\mathcal H_L\), recover eqn.~(2.1) of~\cite{ASS2023} .
We choose \((- ,+,+)\) for three pages.

As in~\cite{ASS2023}, we assume that the heavy insertion admits a semiclassical
description as a spherical dust shell of total mass \(m\).
The eigenstate thermalization hypothesis (ETH) used in~\cite{ASS2023}
separates the heavy-operator matrix elements into a smooth energy-dependent
envelope and fluctuating coefficients
\cite[Eqs.~(D.2)--(D.3)]{ASS2023}, following ref.~\cite{Sasieta:2022ksu}.
We assume the corresponding form for the multilinear coefficients
\(\cV_{\boldsymbol n}\), with one energy argument for each page.
Fix finite energy windows \(\mathcal W_i\) with excitation energies above
the vacuum satisfying
\(\ell(E_{i,n_i}-E_{i,\mathrm{vac}})=O(N_c^0)\), and write
\(\mathcal W=\prod_i\mathcal W_i\) for their product.
In the semiclassical large-mass regime of
section~\ref{sec:booklet-saddle}, we take, for
\(\boldsymbol n\in\mathcal W\),
\begin{equation}
 \cV_{\boldsymbol n}\simeq
 \exp\!\left[-\frac12\left(I_0+
             \sum_i\Delta\tau_i E_{i,n_i}\right)\right]R_{\boldsymbol n}.
 \label{eq:sg-heavy-junction-ansatz}
\end{equation}
The energy-independent constant \(I_0\) is the renormalized shell
contribution, and \(\Delta\tau_i>0\) is the shell's Euclidean flight time
on page \(i\), determined by the trajectory below.
Here \(\simeq\) denotes the assumed leading semiclassical envelope,
not a quantitative error bound for individual coefficients.
We take the \(R_{\boldsymbol n}\) to be independent circular complex Gaussians
of zero mean and unit variance, so that
\(\mathbb E(R_{\boldsymbol n}R_{\boldsymbol m}^*)=
\delta_{\boldsymbol n\boldsymbol m}\) and
\(\mathbb E(R_{\boldsymbol n}R_{\boldsymbol m})=0\).
This statistical assumption supplies the Wick contractions used in the
tensor model below.  It does not follow from the large shell mass alone.

Combining the envelope with the Euclidean evolution in
\eqref{eq:sg-booklet-pets} gives energy weights with
\(\beta_i=\widetilde\beta_i+\Delta\tau_i\).
For three pages, let \(\lvert\Psi_{3,\mathcal W}\rangle\) denote the
normalized projection of the multi-PETS onto \(\mathcal W\), assumed nonzero.
The ansatz gives
\begin{equation}
 |\Psi_{3,\mathcal W}\rangle\simeq\frac{1}{\sqrt{\mathcal N_3}}
 \sum_{\boldsymbol n\in\mathcal W}R_{n_1n_2n_3}
 \prod_{i=1}^3 e^{-\beta_iE_{i,n_i}/2}
 |n_1,n_2,n_3\rangle .
 \label{eq:sg-three-leg-eth-state}
\end{equation}
The basis orientations are unchanged, and \(\mathcal N_3\) is the
squared norm of the unnormalized sum.  The common factor \(e^{-I_0/2}\)
has canceled.  This state approximates the full PETS only if the
omitted part has negligible norm.  In tensor-network language,
\(R_{n_1n_2n_3}\) is one
common three-leg tensor, and \(e^{-\beta_iH_i/2}\) supplies the thermal
weight on leg \(i\), with the conjugate-space convention when needed.
The thermal weights factorize by page, but the coefficients still form
one joint tensor rather than three independent pairwise insertions.
The same construction applies to any fixed finite \(\Npg\).
It neither makes a fixed state's marginals exactly thermal nor makes a
thermally weighted state Haar distributed.

\subsection{Euclidean geometry and Lorentzian cosmology}
\label{sec:booklet-saddle}

We use the page-sum Einstein--GHY action and multiway junction condition
of~\cite[section~IV.B, Eqs.~(68)--(76)]{ShenPengLi2024},
with cosmological constant \(-d(d-1)/(2\ell^2)\), bulk matter, and one
dust shell of total mass \(m\).  No additional intrinsic spine
interactions are included.  The asymptotic preparation lengths are fixed,
and endpoint terms and counterterms use one common prescription,
following the page-sum action above and the common subtraction used in~\cite[appendix~A]{ASS2023}.
This choice defines the branched effective theory used here.

The low-temperature candidate consists of thermal-AdS pages with metrics
\begin{equation}
 \mathrm ds_i^2=f_i(r)\,\mathrm d\tau_i^2
 +\frac{\mathrm dr^2}{f_i(r)}+r^2\mathrm d\Omega_{d-1}^2,
 \qquad f_i(r)=f(r)=1+\frac{r^2}{\ell^2}.
 \label{eq:sg-euclidean-page-metric}
\end{equation}
On page \(i\), the shell follows \(r=R_E(T_E)\) and
\(\tau_i=\tau_i(T_E)\), where \(T_E\) is proper Euclidean time.
Its embedding \(\iota_i:W\to V_i\) must induce the same metric from every
page \cite[Eq.~(37)]{ShenPengLi2024},
\begin{equation}
 \begin{gathered}
 \iota_1^*g_1=\cdots=\iota_{\Npg}^*g_{\Npg}
 =h=\mathrm dT_E^2+R_E(T_E)^2\mathrm d\Omega_{d-1}^2,\\
 f_i(R_E)\dot\tau_i^{\,2}
 +\frac{\dot R_E^{\,2}}{f_i(R_E)}=1 .
 \end{gathered}
 \label{eq:sg-common-induced-metric}
\end{equation}
Dots denote derivatives with respect to \(T_E\).

The geometric and Hilbert-space orientations must describe the same
preparation.  Set \(\varsigma_i=\operatorname{sign}\dot\tau_i\).
For the chosen contour we take
\(\boldsymbol\epsilon=\boldsymbol\varsigma=(-,+,\ldots,+)\), giving
\((- ,+)\) for two pages and \((- ,+,+)\) for three.
These signs specify the direction in which the shell traverses each
thermal circle and the corresponding conjugate basis in
\eqref{eq:sg-booklet-pets}.
To fix the extrinsic-curvature signs, we always take the normal into
the retained part of each page.

More explicitly, put the turning point at \(T_E=0\) and
\(\tau_i^c=\tau_i(0)\).  On the circle of period \(\beta_i\), the two
shell endpoints lie at \(\tau_i^c\pm\Delta\tau_i/2\).
We remove the cap between the shell and the short boundary arc of length
\(\Delta\tau_i\).  The retained boundary arc has length
\(\widetilde\beta_i=\beta_i-\Delta\tau_i\).
Writing \(W_i\) for the retained shell face, the maps
\(\varphi_i:W_i\to W\) identify equal \(T_E\) and equal angular
coordinates.  This prescription joins the endpoints with \(T_E\to+\infty\)
to one common insertion, and those with \(T_E\to-\infty\) to its adjoint.
Reversing all orientations interchanges ket and bra.
It also specifies the retained side needed to fix the normals.

Let \(n_i^\mu\) be these inward unit normals and
\(K_{ab}^{(i)}=h_a{}^\mu h_b{}^\nu\nabla_\mu n_{i,\nu}\).
With \(K_i=h^{ab}K_{ab}^{(i)}\), the junction condition is
\begin{equation}
 \sum_{i=1}^{\Npg}\bigl(K_{ab}^{(i)}-K_i h_{ab}\bigr)
 =-8\pi\GN S_{ab}.
 \label{eq:sg-multiway-junction}
\end{equation}
No further orientation sign multiplies an extrinsic curvature already
defined with its inward normal.  The Euclidean convention
\(S_{ab}=2(\sqrt h)^{-1}\delta I_{\rm shell}/\delta h^{ab}\)
gives \(S^a{}_b=\operatorname{diag}(-\sigma,0,\ldots,0)\) for dust, with
\(\sigma=m/(\Omega_{d-1}R_E^{d-1})\).
In particular, the negative entry reflects this Euclidean variation
convention, not a negative shell mass.
There is no bulk energy flux into the shell, so \(m\) is conserved.
We assume stationarity under normal deformations of the junction.
This does not establish stability.

For this retained side,
\(K^\theta{}_{\theta,(i)}
=-\sqrt{f_i(R_E)-\dot R_E^{\,2}}/R_E\).
The temporal component of \eqref{eq:sg-multiway-junction} therefore gives
\begin{equation}
 \sum_{i=1}^{\Npg}\sqrt{f_i(R_E)-\dot R_E^{\,2}}
 =\frac{8\pi\GN m}{(d-1)\Omega_{d-1}R_E^{d-2}} .
 \label{eq:sg-general-unsquared-junction}
\end{equation}
The signs of the roots are fixed before squaring.
For identical thermal-AdS pages, define
\(\kappa_{\Npg}=8\pi\GN m/[\Npg(d-1)\Omega_{d-1}]\).
The radial equation then has the form
\begin{equation}
 \dot R_E^{\,2}+V_E(R_E)=0,\qquad
 V_E(R)=-f(R)+\frac{\kappa_{\Npg}^2}{R^{2d-4}} .
 \label{eq:sg-euclidean-potential}
\end{equation}
The shell travels from infinity to the turning radius
\(R_*^{\,2d-4}f(R_*)=\kappa_{\Npg}^2\) and returns to infinity.
There is one positive turning radius for \(d>2\).
For \(d=2\), it exists when \(4\GN m>\Npg\).
The two-page value
\(\kappa_2=4\pi\GN m/[(d-1)\Omega_{d-1}]\) reproduces the orbit in~\cite{ASS2023}.
For three pages, \(\kappa_3=2\kappa_2/3\), giving
\(R_{*,3}^{\,2d-4}f(R_{*,3})=4\kappa_2^2/9\).

The large-radius regime is measured by
\begin{equation}
 \lambda_{\Npg}:=\frac{\kappa_{\Npg}}{\ell^{d-2}}\gg1,
 \qquad
 \chi:=\frac{R_*}{\ell}\simeq\lambda_{\Npg}^{1/(d-1)} .
 \label{eq:sg-dimensionless-shell-parameter}
\end{equation}
Scaling \(m\ell\) as \(\Nc^2\) makes the shell backreaction classical,
but need not make \(\lambda_{\Npg}\) large.
The common flight time follows by integrating
\(|\dot\tau_i|=\kappa_{\Npg}/[f(R_E)R_E^{d-2}]\) along the orbit,
\begin{equation}
 \Delta\tau_{\Npg}
 =2\int_{R_*}^{\infty}
 \frac{\kappa_{\Npg}\,\mathrm dR}
 {f(R)R^{d-2}\sqrt{-V_E(R)}} .
 \label{eq:sg-euclidean-flight-time}
\end{equation}
This is simply Eq.~(2.12) of~\cite{ASS2023} with \(\kappa_2\) replaced by
\(\kappa_{\Npg}\).  It gives
\begin{equation}
 \beta_i=\widetilde\beta_i+\Delta\tau_{\Npg},\qquad
 \Delta\tau_{\Npg}=O(\ell/\chi)\longrightarrow0
 \quad(\lambda_{\Npg}\to\infty).
 \label{eq:sg-beta-relation}
\end{equation}
Thus the bulk periods approach the boundary preparation lengths in
the large-mass limit used here.

The reflection slice intersects page \(i\) in two components.
At \(\tau_i=\tau_i^c+\beta_i/2\), it contains a complete exterior
hyperbolic slice \(a_i\simeq H^d\).
At \(\tau_i=\tau_i^c\), it contains the central disk
\(c_i=\{0\le r\le R_*\}\).
Only the disk boundaries are identified at
\(w=W\cap\{T_E=0\}\), so the full initial slice is
\begin{equation}
 \Sigma=\left(\bigsqcup_i a_i\right)\sqcup\Sigma_c^{(\Npg)},\qquad
 \Sigma_c^{(\Npg)}
 =\left(\bigsqcup_i c_i\right)\big/
   \bigl(\partial c_i\sim\partial c_j\bigr).
 \label{eq:sg-central-slice}
\end{equation}
The identified boundaries carry the same angular coordinates.
Two disks give the spatial topology \(S^d\).
Three or more give a compact branched space, with no ordinary manifold
neighborhood at \(w\) \cite[Appendix~C]{ShenPengLi2024}.
We also write \(c^{(\Npg)}=\Sigma_c^{(\Npg)}\) when labeling a bulk
subsystem.  This global construction, rather than the radial equation
alone, explains the separate exteriors and the closed center.

\paragraph{Lorentzian continuation}
Continue about the reflection slice by
\(T_E=\mathrm iT\) and
\(\tau_i-\tau_i^c=\mathrm i\varsigma_i t_i\), where \(t_i\) is
future-directed static AdS time in the central patch.
Assuming compatible continuation of the face identifications, the exteriors
remain separate and the central patches still meet only at the shell.
Writing dots now for \(\mathrm d/\mathrm dT\), the shell equations become
\begin{align}
 \dot R^{\,2}+V_L(R)&=0,\qquad V_L(R)=-V_E(R),
 \label{eq:sg-lorentzian-potential}\\
 \ddot R&=-\frac{R}{\ell^2}
 -\frac{(d-2)\kappa_{\Npg}^2}{R^{2d-3}} .
 \label{eq:sg-lorentzian-second-order}
\end{align}
The initial conditions are \(R(0)=R_*\) and \(\dot R(0)=0\).
The shell therefore contracts toward a formal big crunch, with a
time-reversed big bang in the past.
Its proper time to collapse is
\begin{equation}
 T_f^{(\Npg)}=\int_0^{R_*}\frac{\mathrm dR}{\sqrt{-V_L(R)}}
 =\frac{\pi\ell}{2(d-1)}+O(\ell\chi^{-2}),\qquad d>2 .
 \label{eq:sg-proper-collapse-time}
\end{equation}
For \(d=2\), \(T_f^{(\Npg)}=\pi\ell/2\) exactly.
The leading coefficient follows by rescaling \(R=R_*x\), which gives
\(\ell\int_0^1 x^{d-2}\mathrm dx/\sqrt{1-x^{2d-2}}\).
For three pages the same formula holds with
\(\chi=\chi_3:=R_{*,3}/\ell\).

The static time along the shell satisfies
\(\mathrm dt_i/\mathrm dT=\kappa_{\Npg}/[f(R)R^{d-2}]>0\),
with \(t_i(0)=0\).  Its endpoint \(t_f^{(\Npg)}\) approaches
\(\pi\ell/2\) as \(\chi\to\infty\), and is distinct from the proper time
\(T_f^{(\Npg)}\).
Each locally AdS central patch also admits an open-FLRW chart,
\begin{equation}
 \mathrm ds_i^2=-\mathrm dt_{c,i}^2
 +\ell^2\cos^2(t_{c,i}/\ell)\,\mathrm dH_d^2,
 \label{eq:sg-pagewise-flrw}
\end{equation}
where \(t_{c,i}\) is cosmological time and \(\mathrm dH_d^2\) is the unit
hyperbolic metric.
For the selected shell branch, this chart covers each central patch up to
the excluded singular endpoints.

\subsection{Bulk state and probe excitations}
\label{sec:bulk-state}

The same Euclidean preparation defines a pure state of the bulk fields
on \(\Sigma\) in
\eqref{eq:sg-central-slice}.  We denote it by \(\lvert\psi_0\rangle\) without light
insertions and by \(\lvert\psi_{\mathcal I}\rangle\) with insertion lists
\(\mathcal I=(\mathcal I_1,\ldots,\mathcal I_{\Npg})\).
Each \(\mathcal I_i\) specifies the light fields, positions and order on
the ket preparation contour of page \(i\).
The bra contains their reflected adjoints in reverse order.
We assume that the chosen bulk theory and its matching conditions at \(W\)
admit this preparation, using the same regulator in all subsequent sewing. 

At fixed regulator, we describe the fields in the kinematic space
\(\bigotimes_i(\mathcal H_{a_i}^{\rm reg}\otimes
\mathcal H_{c_i}^{\rm reg})\).
The factors \(c_i\) refer to the page patches in
\eqref{eq:sg-central-slice}, not independent closed universes.
Physical states obey the spine matching and Gauss constraints within this
description.  Partial traces and entropies below use this fixed
subregion convention.  As in~\cite{ASS2023}, we assume that the resulting bulk state
can be approximated by exterior--central pure pairs at large \(\lambda_{\Npg}\) limit,
\begin{equation}
 \lvert\psi_{\mathcal I}\rangle
 \simeq\bigotimes_{i=1}^{\Npg}
 \lvert\Phi_i[\mathcal I_i]\rangle_{a_i c_i}.
 \label{eq:sg-bulk-factorization}
\end{equation}
In this case, \(\Delta\tau_{\Npg}\rightarrow 0 \) such that the cosmology parts decouple from each other and that the bulk state factorizes.
It requires \(\lambda_{\Npg}\gg1\) and pair entanglement large compared
with residual correlations between central patches, including those
required by matching.

To relate these states to the replica calculation, let
\(\rho_X\) denote the reduction of
\(\lvert\psi_{\mathcal I}\rangle\) to a bulk region \(X\).
In the pair approximation, tracing out \(c_i\) sews the ket and reflected
bra into a page contour of total length \(\beta_i\).
Let \(H_{b,i}\) be the static bulk Hamiltonian on the exterior factor
\(\mathcal H_{a_i}^{\rm reg}\).
Write \(\mathcal I_i^{\rm pair}\) for the ket insertions together with
their reflected bra partners, and let \(\mathcal T_{E,i}\) order them
along the oriented contour, with later times to the left.
The positions \(x_{i,\alpha}\) include Euclidean time and spatial
coordinates, and \(\phi^E_{i,\alpha}\) are Euclidean Heisenberg fields.
The expressions (2.26)--(2.27) of \cite{ASS2023} then generalize to
\begin{align}
 \widehat\varrho_i[\mathcal I_i]
 &\simeq e^{-\beta_iH_{b,i}}
  \mathcal T_{E,i}\!\left\{
  \prod_{\alpha\in\mathcal I_i^{\rm pair}}
  \phi^E_{i,\alpha}(x_{i,\alpha})\right\},\nonumber\\
 Z_i[\mathcal I_i]&:=\Tr\widehat\varrho_i[\mathcal I_i]>0,\qquad
 \varrho_i[\mathcal I_i]
 :=\frac{\widehat\varrho_i[\mathcal I_i]}{Z_i[\mathcal I_i]} .
 \label{eq:sg-euclidean-page-state}
\end{align}
Here \(\widehat\varrho_i\) is the positive operator obtained by that
sewing, and \(\rho_{a_i}\simeq\varrho_i\) within the approximation.
Equivalently, if \(B_i\) is the coefficient matrix of the unnormalized
pair state, the sewn operator is \(B_iB_i^\dagger\).
Without insertions, the right-hand side is simply \(e^{-\beta_iH_{b,i}}\).
These are bulk-field density operators, distinct from the exact boundary
marginals of \eqref{eq:sg-booklet-pets}.

For a page subset \(A\), define \(a_A=\bigsqcup_{i\in A}a_i\) and
\(S_i=S(\rho_{a_i})\), where \(S(\rho)=-\Tr(\rho\log\rho)\).
Pure pairs have equal spectra on their two factors, so
\begin{equation}
 S_{\rm bulk}(a_A)\simeq\sum_{i\in A}S_i,\qquad
 S_{\rm bulk}(c^{(\Npg)})\simeq\sum_{i=1}^{\Npg}S_i .
 \label{eq:sg-actual-bulk-reductions}
\end{equation}
Thus the total bulk state is pure while its central reduction is
generally mixed.  These entropy sums are the leading pair approximation
used in section~\ref{sec:replica-island}.
Residual correlations that change the sums invalidate that simplification.

Following~\cite[section~2.2]{ASS2023}, we consider light
insertions whose combined backreaction on the page geometries
and the shell trajectory can be neglected. We evaluate the
corresponding bulk-field path integrals on the background
constructed in section~\ref{sec:booklet-saddle}, within the
semiclassical regime specified there.

% ===== END sections/states_geometry =====
%
% ===== BEGIN sections/replica_island =====
\section{Replica moments and entanglement in the booklet cosmology state}
\label{sec:replica-island}
In this section we calculate the entanglement wedge of the boundaries through the replica method of \cite{ASS2023}.  The two candidate
bulk regions contain either the corresponding exteriors alone or those
exteriors together with the entire cosmology.  Their competition gives a
conditional island formula, provided the large-mass factorization and
the saddle-selection assumptions below hold.

Let $A\subseteq\{1,\ldots,\Npg\}$ label the boundary CFTs whose entropy
we seek, and let $\bar A$ contain the remaining pages.  For an integer
$n\geq2$, the $n$th moment of their reduced density matrix is
\begin{equation}
 \begin{aligned}
 \widehat\rho_A&:=\Tr_{\bar A}|\widehat\Psi_{\cV}\rangle
 \langle\widehat\Psi_{\cV}|,\qquad
 \rho_A:=\frac{\widehat\rho_A}{Z_{\cV}},\\
 \Tr\rho_A^n&=\frac{\Tr\widehat\rho_A^n}{Z_{\cV}^n}.
 \end{aligned}
 \label{eq:exact-normalized-moment}
\end{equation}
Here $|\widehat\Psi_{\cV}\rangle$ is the unnormalized boundary state of
section~\ref{sec:booklet-state}, and
$Z_{\cV}=\langle\widehat\Psi_{\cV}|\widehat\Psi_{\cV}\rangle$ is its norm.
Equation~\eqref{eq:exact-normalized-moment} is an exact boundary identity.
To evaluate this ratio semiclassically, we take \(n\) copies of the
Euclidean preparation.  For the numerator, the boundary cuts are
glued cyclically between replicas on the pages in \(A\), while those
on the remaining pages are glued within each replica.  For the
denominator, all boundary cuts are glued within each replica, as
required for \(Z_{\mathcal V}^{\,n}\).  We evaluate each gravitational
path integral by summing over the contributing bulk saddles and
denote the ratio of these two sums by
\(\mathcal M_{A,\mathrm{grav}}(n)\).  Bulk saddles connecting different
replicas can contribute to either sum, including the normalization.
Following \cite{ASS2023}, we therefore use
\(\Tr\rho_A^n\simeq\mathcal M_{A,\mathrm{grav}}(n)\)
under the semiclassical assumptions stated below..

We begin with the bulk reference state $|\psi_0\rangle$, prepared without
light insertions.  We restrict attention to reference states for which
the pair approximation \eqref{eq:sg-bulk-factorization} is valid.
In particular, the pair entanglement must dominate residual correlations
between central patches.  Large shell mass alone does not guarantee this
hierarchy for an unexcited state below the bulk energy gap.

Write $\rho_{a_i}$ for the normalized reduction of $|\psi_0\rangle$
to bulk exterior $a_i$, suppressing the empty insertion list
$\varnothing$.  The one-boundary CFT state is instead denoted by
$\rho_{\{i\}}$.
The Euclidean operator $\widehat\varrho_i[\varnothing]$ in
\eqref{eq:sg-euclidean-page-state} acts on
$\mathcal H_{a_i}^{\rm reg}$, as does the exterior bulk Hamiltonian
$H_{b,i}$.  The normalization
$Z_i:=\Tr\widehat\varrho_i[\varnothing]>0$ is a bulk page normalization,
not the boundary norm $Z_{\cV}$.
In the factorized replica geometry, a closed loop through $k$ copies
of page $i$ sews $k$ such operators together.  For a positive integer
$k$, we use the leading loop weight
\begin{equation}
 \begin{aligned}
 q_i(k)&:=\Tr\rho_{a_i}^k,\\
 \mathcal Z_i(k)&:=Z_i^kq_i(k)
 \simeq\Tr\!\left[\widehat\varrho_i[\varnothing]^k\right].
 \end{aligned}
 \label{eq:page-loop-weight}
\end{equation}
All traces in this equation are over $\mathcal H_{a_i}^{\rm reg}$.
The approximation uses
$\rho_{a_i}\simeq\widehat\varrho_i[\varnothing]/Z_i$, rather than an exact
identification of the actual bulk reduction with the Euclidean operator.
Thus $q_i(1)=1$, while $q_i(2)$ is the exterior purity.
For a set of pages, define
$q_A(k):=\prod_{i\in A}q_i(k)$ and
$q_{\bar A}(k):=\prod_{i\in\bar A}q_i(k)$, with empty products equal to one.
These are products of bulk page moments, not moments of the boundary
density matrix $\rho_A$.

\subsection{Purity and replica moments}
\label{sec:purity}
\label{sec:integer-replicas}

For two replicas, the numerator of \eqref{eq:exact-normalized-moment}
glues the ket cut in copy 1 to the bra cut in copy 2 on each boundary
in $A$, and conversely.  On $\bar A$, each ket cut is joined to the
bra cut in the same copy.  The normalization instead uses this latter,
identity gluing on every boundary.

The shell connections in the interior are a separate choice.
Identity pairing $e$ joins each ket junction to the bra junction in the
same replica.  Swap pairing $\eta_2=(12)$ joins it to the bra junction
in the other replica, giving the replica-connected saddle.
Each shell carries all $\Npg$ page faces into the same common spine,
with the junction conditions of section~\ref{sec:states-geometry}.
Changing its replica pairing therefore reconnects all pages together,
not one page at a time.

Denote a saddle contribution by $\mathfrak Z_{2,A}[g]$, where $A$
specifies the boundary gluing and $g=e,\eta_2$ specifies the shell
pairing.  The subscript $\varnothing$ means identity boundary gluing
on every page and hence a contribution to the normalization.
Identity pairing in the numerator gives one two-copy loop on each
page in $A$ and two one-copy loops on each page in $\bar A$.
Swap pairing exchanges these roles.  In the denominator, identity
pairing gives only one-copy loops, while swap pairing gives a
two-copy loop on every page.

We adopt the same large-mass factorization as in~\cite[eqs.~(3.5)--(3.8)]{ASS2023} for these four geometries.
After the page traces are separated, let $\mathcal W_2$ denote their
common remaining gravitational weight, including shell and spine terms.
Its independence of the pairing is an assumption about the selected
branched saddles.  The resulting contributions are
\begin{equation}
 \begin{pmatrix}
 \mathfrak Z_{2,A}[e]&\mathfrak Z_{2,A}[\eta_2]\\
 \mathfrak Z_{2,\varnothing}[e]&\mathfrak Z_{2,\varnothing}[\eta_2]
 \end{pmatrix}
 \simeq\mathcal W_2\prod_iZ_i^2
 \begin{pmatrix}
 q_A(2)&q_{\bar A}(2)\\
 1&q_A(2)q_{\bar A}(2)
 \end{pmatrix}.
 \label{eq:four-purity-weights}
\end{equation}
Assuming that these are all contributions at the retained order, we
sum the first row and divide by the sum of the second.
The common factor cancels, giving
\begin{equation}
 \mathcal M_{A,\mathrm{grav}}(2)
 \simeq\frac{q_A(2)+q_{\bar A}(2)}
 {1+q_A(2)q_{\bar A}(2)} .
 \label{eq:booklet-purity}
\end{equation}
Keeping the connected normalization term makes this expression equal
to one for empty or full $A$, as required for a pure boundary state.
In the bulk pair approximation, the two numerator terms are the purities
of the exterior unions $a_A=\bigsqcup_{i\in A}a_i$ and $a_{\bar A}$.
Purity of the full bulk state also identifies the second with the purity
of $a_A$ together with the entire cosmology.  This is the origin of the two
island candidates, rather than an independent assumption about the entropy.
For three pages, the purity of one boundary equals that of its complement,
\begin{equation}
 \mathcal M_{\{1\},\mathrm{grav}}(2)
 \simeq\mathcal M_{\{2,3\},\mathrm{grav}}(2)
 \simeq
 \frac{q_1(2)+q_2(2)q_3(2)}
 {1+q_1(2)q_2(2)q_3(2)}.
 \label{eq:purity-three-page}
\end{equation}

For general integer $n$, label replicas by $r=1,\ldots,n$.
Let $S_n$ be their permutation group, $e$ its identity, and
$\eta_n=(1\,2\cdots n)$ the cyclic permutation.
The boundary gluing is specified on each page by
\begin{equation}
 \sigma_i^A=
 \begin{cases}
  \eta_n,&i\in A,\\
  e,&i\in\bar A.
 \end{cases}
 \label{eq:boundary-gluing-permutation}
\end{equation}
One permutation $g\in S_n$ pairs ket junction $r$ with bra junction
$g(r)$ on every page.  Passing through this pairing and then the
boundary gluing returns a page leg to replica $\sigma_i^Ag(r)$.
Consequently, each cycle of $\sigma_i^Ag$ forms one closed page loop.
We compose permutations from right to left and write
$\operatorname{cyc}(p)$ for the list of cycle lengths of a permutation
$p$, including repeated lengths.  In the normalization,
$\sigma_i^\varnothing=e$, so the loops are the cycles of $g$.

Let $\mathfrak Z_{n,A}[g]$ denote the corresponding $n$-replica saddle
contribution, extending the two-replica notation above.
At each fixed $n$, we assume that these saddles exist, exhaust the
retained order and have a common gravitational factor independent of
$A$ and $g$.  Assigning the weight \eqref{eq:page-loop-weight} to every
loop then gives
\begin{equation}
 \begin{aligned}
 \mathcal M_{A,\mathrm{grav}}(n)
 &:=\frac{\sum_{g\in S_n}\mathfrak Z_{n,A}[g]}
              {\sum_{g\in S_n}\mathfrak Z_{n,\varnothing}[g]}\\
 &\simeq
 \frac{\displaystyle\sum_{g\in S_n}
  \prod_{i=1}^{\Npg}
  \prod_{k\in\operatorname{cyc}(\sigma_i^Ag)}\mathcal Z_i(k)}
 {\displaystyle\sum_{g\in S_n}
  \prod_{i=1}^{\Npg}
  \prod_{k\in\operatorname{cyc}(g)}\mathcal Z_i(k)} .
 \end{aligned}
 \label{eq:gravity-cycle-ratio}
\end{equation}
The factors $Z_i^n$ cancel because the cycle lengths on each page
sum to $n$.  Additional pairing-dependent shell or spine terms would
change the ratio.
For two pages the result reduces to eq.~(3.13) in~\cite{ASS2023}.

For replicas of a single probe-excited state, the insertion list
$\mathcal I_i$ is the same in every copy of page $i$.
The operator $\widehat\varrho_i[\mathcal I_i]$ includes the ordered ket
insertions and their reflected bra partners, as in
section~\ref{sec:bulk-state}.  A loop through $k$ copies therefore has
weight
\begin{equation}
 \mathcal Z_i(k;\mathcal I_i)
 \simeq\Tr\!\left[\widehat\varrho_i[\mathcal I_i]^k\right].
 \label{eq:paper-probe-inserted-trace}
\end{equation}
This is the pagewise generalization of ~\cite[eq.~(3.9)]{ASS2023}, within the same large-mass approximation.
Numerator and normalization must use these same insertions.
The ordering and backreaction restrictions of
section~\ref{sec:bulk-state} still apply.
Different insertion lists in different replicas would require keeping
track of the ket--bra pairings, rather than using this single-state trace.
\subsection{The whole-cosmology island}
\label{sec:island}

We return to the reference state and ask which bulk region determines
the boundary von Neumann entropy.  For the exact density matrix,
\begin{equation}
 S(\rho_A):=-\Tr(\rho_A\log\rho_A)
 =-\left.\partial_n\log\Tr\rho_A^n\right|_{n=1}.
 \label{eq:boundary-replica-entropy}
\end{equation}
Using this identity in gravity requires a prescription for continuation
away from integer $n$.  Following~\cite[section~3.3, eq.~(3.15)]{ASS2023}, we retain the replica-symmetric
families $g=e$ and $g=\eta_n^{-1}$.
Their quotients identify points related by the cyclic relabeling of
replicas, including the corresponding copies of the common spine.
It is these quotient geometries that are continued in $n$, not the
permutations themselves.  We assume that the continuation preserves the
selected branched junction conditions and that both families remain
allowed and stable, while other saddles are subleading.

The identity pairing has an $n$-cycle on pages in $A$ and one-cycles on
$\bar A$, giving numerator weight $q_A(n)$ after normalization by
$\prod_i Z_i^n$.  The pairing $\eta_n^{-1}$ reverses these roles and
gives $q_{\bar A}(n)$.
Their normalization weights are $1$ and $q_A(n)q_{\bar A}(n)$.
Denoting the ratio restricted to these two saddles by
$\mathcal M_A^{(2\mathrm{s})}(n)$, we obtain
\begin{equation}
 \mathcal M_A^{(2\mathrm{s})}(n)
 \simeq\frac{q_A(n)+q_{\bar A}(n)}{1+q_A(n)q_{\bar A}(n)}.
 \label{eq:main-two-saddle-continuation}
\end{equation}
This expression displays the branch competition.  It is not an exact
finite-parameter continuation to which one can directly apply
\eqref{eq:boundary-replica-entropy}.

Define the bulk entropy sums
$s_A:=\sum_{i\in A}S(\rho_{a_i})$ and
$s_{\bar A}:=\sum_{i\in\bar A}S(\rho_{a_i})$.
To select a sharp branch, we assume a family of reference states with
large entropy sum and entropy difference.
For $s_A<s_{\bar A}$, the required spectral hierarchy in a range
$0<n-1\ll1$ is
\begin{equation}
 \frac{q_{\bar A}(n)}{q_A(n)}\ll1,\qquad
 q_A(n)q_{\bar A}(n)\ll1 .
 \label{eq:main-branch-selection-window}
\end{equation}
The first inequality selects the numerator branch and the second
suppresses the connected normalization.
The reversed entropy ordering requires the reversed ratio.
The associated entropy window for this branch-selection prescription is
\begin{equation}
 (n-1)|s_A-s_{\bar A}|\gg1,\qquad
 (n-1)(s_A+s_{\bar A})\gg1 .
 \label{eq:main-entropy-selection-window}
\end{equation}
Those inequalities alone do not guarantee
\eqref{eq:main-branch-selection-window} for arbitrary spectra.
The reference-state family must also have compatible R\'enyi ordering.
We assume such a family rather than deriving it for every unexcited
low-temperature preparation.

As in \cite{ASS2023}, We select the dominant branch in
this regime and then continue that branch to $n=1$.
For $s_A<s_{\bar A}$ it is $q_A(n)$, whose entropy is
\begin{equation}
 -\left.\partial_n\log q_A(n)\right|_{n=1}
 =\sum_{i\in A}S(\rho_{a_i})=s_A.
 \label{eq:replica-selected-branch-entropy}
\end{equation}
The opposite ordering gives $s_{\bar A}$.
At fixed finite parameters, differentiating the full ratio
\eqref{eq:main-two-saddle-continuation} would instead cancel numerator
and denominator slopes at $n=1$.  The minimum obtained in this order is
therefore a semiclassical result, not an exact entropy identity for
the finite ratio.

The geometric interpretation follows from purity of the full bulk state.
Recall that $c^{(\Npg)}$ is the entire closed cosmology, formed by joining
the patches $c_i$ along the spatial spine $w=W\cap\Sigma$.
Here $W$ is the spacetime junction and $\Sigma$ the reflection slice.
Bulk reductions use the regulated factors of
section~\ref{sec:bulk-state}.
Write $S_{\rm bulk}(X)$ for the entropy of the bulk state reduced to
region $X$, and use $\sqcup$ to join disconnected components.
The complement of $a_{\bar A}$ is $a_A\sqcup c^{(\Npg)}$, so
\begin{equation}
 S_{\rm bulk}(a_A\sqcup c^{(\Npg)})=S_{\rm bulk}(a_{\bar A}).
 \label{eq:island-regions}
\end{equation}
Adding the closed cosmology changes neither the asymptotic boundary nor
the internal boundary of the candidate region.  Its spine is an
internal junction, not a cut through the center.
Both candidates thus have an empty extremal surface.
With the common spine contribution assumed above, there is no
additional relative term in this two-saddle comparison.
The resulting island formula is 
\begin{equation}
 \begin{aligned}
 S(\rho_A)
 &\simeq\min\{s_A,s_{\bar A}\}\\
 &\simeq\min\!\left\{
 S_{\rm bulk}(a_A),\,
 S_{\rm bulk}(a_A\sqcup c^{(\Npg)})\right\}.
 \end{aligned}
 \label{eq:branched-center-island-formula}
\end{equation}
The second line uses entropy additivity in the pair approximation.
Only the exterior-only and whole-center candidates have been compared.
The sharp minimum is not controlled near $s_A=s_{\bar A}$, or when
other saddles or leading interpage correlations invalidate the assumptions.

For three pages with equal bulk exterior entropies
$S(\rho_{a_i})=s>0$, in the regime where branch selection is controlled,
\begin{equation}
 S(\rho_A)
 \simeq\min\{|A|,3-|A|\}\,s .
 \label{eq:three-page-island-check}
\end{equation}
Here $|A|$ counts the retained pages.
One page selects its exterior alone, whereas two pages select their
exteriors together with the entire center.
In the three-page example, the whole-center island associated with two
retained boundaries motivates joint access to central probes.
It does not establish a recovery channel.  To investigate that access,
we next define the map from central probe states to the boundary.
%
% ===== END sections/replica_island =====
\section{The restricted cosmology-to-boundary map}
\label{sec:boundary-map}

The island calculation concerns the entropy of a reference state. To ask how central excitations are represented at the boundary, we must instead compare a family of states prepared on the same background. Following~\cite[sections~4.1--4.3]{ASS2023}, we define the corresponding
linear map and study its overlaps and rank.  Under the probe and saddle assumptions below, the replica calculation suggests that the linear map is non-isometric and we give the generalization of the results in~\cite{ASS2023}.

\subsection{Probe states and boundary overlaps}
\label{sec:gram-moments}

Fix a bulk reference preparation in the probe regime of
section~\ref{sec:bulk-state}, with light-insertion list $\mathcal I_{\rm ref}$.
Keep its heavy insertion, Euclidean lengths and reference light insertions
unchanged, and vary only probes whose dominant Euclidean worldlines
intersect the cosmological part of the reflection slice.
We choose the probe insertions using the Euclidean geodesic
prescription of \cite{ASS2023}.
In the regime \(1\ll\Delta\ll N_c^2\), the inner product of a boundary insertion and its conjugate are prepared by a dominant geodesic connecting
the two insertions in the Euclidean saddle.  
We select insertions for which this
geodesic crosses the cosmological part of the reflection slice,
where it prepares a localized bulk particle in the worldline
approximation.  Keeping the heavy insertion and the reference
preparation fixed, these additional insertions specify both
the bulk probe state \(|\psi_I\rangle\) and its unnormalized
boundary counterpart \(|\widehat\Psi_I\rangle\).
The map \(V_\psi\) is defined by this correspondence and extended
linearly to the chosen probe space.
This correspondence defines
\begin{equation}
 V_\psi:\mathcal H_\psi\longrightarrow
 \bigotimes_{i=1}^{\Npg}\mathcal H_i^{\rm reg},
 \qquad
 V_\psi|\psi_I\rangle=|\widehat\Psi_I\rangle .
 \label{eq:restricted-euclidean-map}
\end{equation}
We keep the boundary states unnormalized and extend the map
linearly to superpositions.  The map is restricted to the chosen
probe space and need not preserve inner products.

We follow the conventional Euclidean construction but keep the boundary images
unnormalized so that $V_\psi$ is linear.
For individual normalized images, put
$|\Psi_I\rangle=|\widehat\Psi_I\rangle/\sqrt{Z_I}$, where
$Z_I=\|\widehat\Psi_I\|^2>0$.  Their Gram matrix is
\begin{equation}
 G_{IJ}:=\langle\Psi_I|\Psi_J\rangle
 =\frac{\widehat{\mathcal Z}_{IJ}}{\sqrt{Z_I Z_J}},
 \qquad
 \widehat{\mathcal Z}_{IJ}:=
 \langle\widehat\Psi_I|\widehat\Psi_J\rangle .
 \label{eq:normalized-probe-gram}
\end{equation}
Thus $G_{II}=1$ exactly.  This matrix is not $V_\psi^\dagger V_\psi$,
although the two have the same rank.
Normalizing the image of every superposition separately would give
a nonlinear map.

Let $\Delta_{I,\alpha}$ be the conformal dimension of the $\alpha$-th
probe insertion preparing $|\psi_I\rangle$.
We require $1\ll\Delta_{I,\alpha}\ll N_c^2$ for the particle-worldline
approximation, with the combined backreaction of the selected probes
negligible as assumed in section~\ref{sec:bulk-state}.
The single-copy semiclassical saddle gives the smooth overlap
\cite[eq.~(4.11)]{ASS2023},
\begin{equation}
 [G_{IJ}]_{\rm sc}\simeq\delta_{IJ}.
 \label{eq:semiclassical-smooth-gram}
\end{equation}
The brackets denote a coarse-grained saddle estimate.
For distinct states, this result reproduces bulk orthogonality.
It does not determine the second moment of the overlap, which also
receives a connected two-replica contribution.

To evaluate that contribution, let $\rho_i^{\rm ref}$ be the normalized
bulk density matrix on exterior $a_i$ in the reference preparation.
Writing $\mathcal I_{{\rm ref},i}$ for its page-$i$ insertion list,
the pair approximation of section~\ref{sec:bulk-state} gives
\begin{equation}
 \rho_i^{\rm ref}
 \simeq
 \frac{\widehat\varrho_i[\mathcal I_{{\rm ref},i}]}
      {Z_i[\mathcal I_{{\rm ref},i}]},
 \qquad
 \Tr\rho_i^{\rm ref}=1.
 \label{eq:gram-reference-page-state}
\end{equation}
The bulk normalization $Z_i$ is distinct from the boundary norm $Z_I$.
For this reference state, write $q_i(n)=\Tr[(\rho_i^{\rm ref})^n]$.
The loop weight $\mathcal Z_i(n;\mathcal I_{{\rm ref},i})$ of
eq.~\eqref{eq:paper-probe-inserted-trace} sews $n$ copies of the
unnormalized page operator.  Each copy includes the same ordered
reference insertions.  It reduces to the thermal partition function
$\Tr e^{-n\beta_iH_{b,i}}$ only when those insertions are absent.

At large shell mass and large $N_c$, the common shell contribution and the leading type-I worldline factors cancel against the norm
saddles~\cite[eqs.~(4.12)--(4.18)]{ASS2023}.
We also assume that the leading probe fluctuation factors cancel and that the reference page weights remain unchanged.
We then get
\begin{equation}
 \begin{aligned}
 [|G_{IJ}|^2]_{\rm sc,conn}
 &\simeq
 \prod_{i=1}^{\Npg}
 \frac{\mathcal Z_i(2;\mathcal I_{{\rm ref},i})}
      {\mathcal Z_i(1;\mathcal I_{{\rm ref},i})^2}\\
 &\simeq
 \prod_i\Tr[(\rho_i^{\rm ref})^2]
 =\exp\!\left[-\sum_i S_2(\rho_i^{\rm ref})\right],
 \qquad I\ne J,
 \end{aligned}
 \label{eq:gram-connected-variance}
\end{equation}
where $S_2(\rho)=-\log\Tr\rho^2$.
The normalization is already included in each page purity.
Greater second R\'enyi entanglement between the exteriors and the
center therefore suppresses the overlap fluctuations.

At each fixed $2\leq n\leq K_\psi$, impose the same cancellation
assumptions and retain the leading connected cyclic saddle.
For pairwise distinct labels $I_1,\ldots,I_n$, we get
\begin{equation}
 [G_{I_1I_2}G_{I_2I_3}\cdots G_{I_nI_1}]_{\rm sc,conn}
 \simeq
 \prod_i
 \frac{\mathcal Z_i(n;\mathcal I_{{\rm ref},i})}
      {\mathcal Z_i(1;\mathcal I_{{\rm ref},i})^n}
 \simeq\prod_iq_i(n).
 \label{eq:connected-gram-moment}
\end{equation}
These are ratios of separately evaluated replica and norm contributions.
They are not generally exact ensemble identities for individually normalized
Gaussian states.  Appendix~\ref{app:boundary-network-support}
gives the cancellation and explains this distinction.
No probability for the overlaps of a single fixed CFT is inferred.
The nonzero second moment shows why the smooth overlap alone does
not establish isometry.  Nonorthogonal boundary images may still be
linearly independent, which motivates the rank question below.
\subsection{Microcanonical overlaps and rank}
\label{sec:rank-resolvent}

The number of independent boundary images is finite after restricting
each page to a microcanonical window.  Let $P_i$ project onto
$[E_i,E_i+\Delta E_i)$, with dimension $D_i>0$, and write
$P_E=\bigotimes_iP_i$ and $D_\partial=\prod_iD_i$.
The projected map is $V_E^{\rm raw}=P_EV_\psi$, with raw Gram matrix
$\widehat G^E=(V_E^{\rm raw})^\dagger V_E^{\rm raw}$.
For the probe family of section~\ref{sec:gram-moments}, assume a common
leading window probability
$\langle\Psi_I|P_E|\Psi_I\rangle\simeq p_E>0$.
With $(D_Z)_{IJ}=Z_I\delta_{IJ}$, we rescale the overlaps to unit
leading diagonal,
\begin{equation}
 (G_E)_{IJ}
 =\frac{\langle\Psi_I|P_E|\Psi_J\rangle}{p_E},
 \qquad
 G_E=\frac1{p_E}D_Z^{-1/2}\widehat G^E D_Z^{-1/2}.
 \label{eq:microcanonical-map-gram}
\end{equation}
This invertible rescaling does not change the rank.

Following~\cite[eqs.~(4.24)--(4.26)]{ASS2023}, assume
that the reference page weights are full rank and approximately
constant within the chosen windows.  Each window-normalized page cycle then
contributes $D_i^{1-n}$, so eq.~\eqref{eq:connected-gram-moment} becomes
\begin{equation}
 [(G_E)_{I_1I_2}\cdots(G_E)_{I_nI_1}]_{\rm sc,conn}
 \simeq\prod_iD_i^{1-n}=D_\partial^{1-n}
 \label{eq:microcanonical-connected-cycle}
\end{equation}
for distinct labels at fixed $n\geq2$.
As in~\cite{ASS2023}, retain the leading noncrossing contractions of
these cycles as $K_\psi,D_\partial$ grow at fixed positive ratio.
For complex $z$ off the spectrum, let
$R_E(z)=[\Tr(zI-G_E)^{-1}]_{\rm sc}$.
A connected cycle of length $n$ containing a marked point of the
trace leaves $n$ intervals.  Summing their noncrossing insertions gives
$R_E^n$, multiplied by the cycle weight $D_\partial^{1-n}$.
The planar resummation gives~\cite[eqs.~(4.29)--(4.31)]{ASS2023}
\begin{equation}
 zR_E(z)
 =K_\psi+D_\partial\sum_{n=1}^{\infty}
       \left(\frac{R_E(z)}{D_\partial}\right)^n
 =K_\psi+\frac{D_\partial R_E(z)}{D_\partial-R_E(z)}.
 \label{eq:microcanonical-resolvent}
\end{equation}
The series is initially evaluated at large $|z|$.
The branch satisfying $R_E(z)\sim K_\psi/z$ fixes the solution.
Its discontinuity
$\mathcal D_E(\lambda)=-\pi^{-1}\operatorname{Im}R_E(\lambda+i0)$
gives the leading eigenvalue density
\cite[eqs.~(4.32)--(4.33)]{ASS2023},
\begin{equation}
 \mathcal D_E(\lambda)
 =\frac{D_\partial}{2\pi\lambda}
   \sqrt{(\lambda_+-\lambda)(\lambda-\lambda_-)}
   \,\mathbf1_{[\lambda_-,\lambda_+]}(\lambda)
   +(K_\psi-D_\partial)_+\delta(\lambda),~
 \lambda_\pm =\left(1\pm\sqrt{\frac{K_\psi}{D_\partial}}\right)^2.
 \label{eq:microcanonical-eigenvalue-density}
\end{equation}
Here $\mathbf1$ restricts the continuous density to the indicated
interval, $(K_\psi-D_\partial)_+ = (K_\psi-D_\partial)\theta(K_\psi-D_\partial)  $,and $\delta(\lambda)$ is the Dirac delta.
The density counts $K_\psi$ eigenvalues in total.
Its zero-eigenvalue term gives $(K_\psi-D_\partial)_+$ null directions,
so the semiclassical calculation predicts
$\min\{K_\psi,D_\partial\}$ independent images in the window.
Therefore we have
\begin{equation}
 \rank\widehat G^E = \min\{K_\psi,D_\partial\}.
 \label{eq:microcanonical-rank-bound}
\end{equation}
\equationaliases{\label{eq:wishart-rank-saturation}}

The bound concerns a finite energy window, not the full preparation map.
Thermal tails outside that window can distinguish states whose
projected images are dependent, as emphasized in~\cite{ASS2023}.
% ===== END sections/boundary_map =====
%
% ===== BEGIN sections/tensor_network =====
\section{A direct multi-arm tensor-network model}
\label{sec:tensor-network}

The tensor network defined in~\cite{ASS2023} gives a
finite-dimensional model of the cosmology-to-boundary map.  Its booklet
extension attaches one exterior--central entangled pair per page to a
common shell tensor.  With equal pair weights and Gaussian shell
coefficients, the map is Gaussian.  Its typical output states reproduce
the island entropy minimum to leading order at large arm dimensions, while its Gram
matrix determines the rank and overlap fluctuations.  The same map will
be used to study recovery from page subsets in
section~\ref{sec:erasure-qec}.

\subsection{One shell tensor, many central inputs}
\label{sec:common-network}

The entangled pairs model the reference state, while a separate input
labels the central excitation to be encoded.  On page \(i\), choose
\(b_i\)-dimensional exterior and central partner spaces and the fixed pair
\begin{equation}
 |\Phi_i\rangle_{a_ic_i}
 =b_i^{-1/2}\sum_{\alpha_i=1}^{b_i}
 |\alpha_i\rangle_{a_i}|\alpha_i\rangle_{c_i}.
 \label{eq:network-pairs-main}
\end{equation}
Fixed isometries \(U_i^{\rm ext}:\mathbb C^{b_i}\to\mathcal H_i^{\rm reg}\)
embed the exterior arms into regulated boundary spaces of dimensions
\(d_i\geq b_i\).  Write \(B=\prod_i b_i\) for the total arm dimension
and \(L=\bigotimes_iU_i^{\rm ext}\) for this boundary embedding.
As in~\cite{ASS2023}, we introduce an
independent central factor \(c_0\) carrying the
input space \(\mathcal H_\psi\), of dimension \(K_\psi\), with basis
\(\{|I\rangle\}\).  It is initially unentangled with the reference
pairs and labels the variable probe, rather than a page patch such as
\(c_3\).
One shell bra \({}_{\rm cen}\langle O_{\rm sh}|\) contracts \(c_0\) and
all the central partners.  Its components in their product basis are
\(X_{\boldsymbol\alpha I}\), with
\(\boldsymbol\alpha=(\alpha_1,\ldots,\alpha_{\Npg})\).

Contracting this bra with \(|I\rangle_{c_0}\) and the central halves of
the pairs leaves an unnormalized state on the exterior arms.
The isometries then map it to the boundary,
\begin{equation}
 V_O|I\rangle
 =\frac1{\sqrt B}\sum_{\boldsymbol\alpha}
 X_{\boldsymbol\alpha I}
 \bigotimes_{i=1}^{\Npg}U_i^{\rm ext}|\alpha_i\rangle_{a_i}.
 \label{eq:network-gaussian-map}
\end{equation}
The pairs, shell bra and exterior isometries remain fixed as the input
varies and are shared by all replicas.
For the random model, take the shell coefficients
\(X_{\boldsymbol\alpha I}\) to be independent, zero-mean standard
circular complex Gaussian variables, with
\begin{equation}
 \mathbb E_{\rm G}[X_{\boldsymbol\alpha I}X^*_{\boldsymbol\beta J}]
 =\delta_{IJ}\prod_i\delta_{\alpha_i\beta_i},
 \qquad
 \mathbb E_{\rm G}[X_{\boldsymbol\alpha I}X_{\boldsymbol\beta J}]=0.
 \label{eq:network-gaussian-covariance}
\end{equation}
We call this the flat Gaussian model, referring to the equal Schmidt
weights of the pairs.  The independent input factor \(c_0\) and Gaussian
shell give independent matrix entries.  More general central inputs can
give correlated entries, as explained in
appendix~\ref{app:boundary-network-support} using the fixed contraction of~\cite{ASS2023}.

\subsection{Islands, rank and selected-state overlaps}
\label{sec:network-results}

First consider the entanglement between a set of pages and its complement.
Let \(|\chi\rangle\) be a normalized input chosen independently of
the shell tensor.  Normalize its nonzero image as
\(|\Psi_O\rangle=V_O|\chi\rangle/\|V_O|\chi\rangle\|\), and let
\(\rho_A^{\rm TN}=\Tr_{\bar A}|\Psi_O\rangle\langle\Psi_O|\).
The effective dimensions on the two sides are
\(B_A=\prod_{i\in A}b_i\) and \(B_{\bar A}=\prod_{i\notin A}b_i\).
The exterior isometries preserve Schmidt coefficients, so the Schmidt
rank cannot exceed either dimension.  Every such network state therefore
satisfies the analogue of the weak island bound in~\cite{ASS2023} eq.~(5.6),
\begin{equation}
 S(\rho_A^{\rm TN})\leq
 \min\{\log B_A,\log B_{\bar A}\}.
 \label{eq:network-island-bound}
\end{equation}
To follow the replica prescription, average the numerator and normalization separately and retain their planar Wick contractions.
Their ratio \(\mathcal M_{n,A}^{\rm pl}\) is the multi-arm form of eq.~(5.11) in~\cite{ASS2023},
\begin{equation}
 \mathcal M_{n,A}^{\rm pl}
 =\frac{\sum_{s=1}^n\mathrm N(n,s)B_A^{n+1-s}B_{\bar A}^{s}}
 {\sum_{s=1}^n\mathrm N(n,s)B^s},
 \qquad \mathrm N(n,s)=\frac1n\binom ns\binom n{s-1}.
 \label{eq:network-ass-planar-ratio}
\end{equation}
The Narayana number \(\mathrm N(n,s)\) counts the noncrossing contractions
with \(s\) cycles.  In the numerator, these contractions give \(s\)
closed index loops on \(\bar A\) and \(n+1-s\) on \(A\).
Each loop contributes its subsystem dimension, while a normalization
loop contributes \(B\).
The ratio is complement symmetric and equals one for empty
or full \(A\).

The entropy follows from the same replica calculation.
Grouping the pages in \(A\) and \(\bar A\) gives two effective
subsystems of dimensions \(B_A\) and \(B_{\bar A}\).
Equation~\eqref{eq:network-ass-planar-ratio} therefore has the
same form as the two-arm result of ASS.
Following its analytic continuation to \(n=1\)~\cite[eqs.~(5.11)--(5.13)]{ASS2025}, we obtain
\[
 \mathbb E_{\rm G}S(\rho_A^{\rm TN})
 \simeq
 -\left.\partial_n\log\mathcal M_{n,A}^{\rm pl}\right|_{n=1}.
\]
Writing \(d_<:=\min(B_A,B_{\bar A})\) and
\(d_>:=\max(B_A,B_{\bar A})\), the leading large-dimension
result is
\begin{equation}
 \mathbb E_{\rm G}S(\rho_A^{\rm TN})
 \simeq \log d_<-\frac{d_<}{2d_>}.
 \label{eq:network-page-entropy}
\end{equation}
The exact Page--Sen result confirms that the nonnegative entropy
deficit \(\log d_<-S(\rho_A^{\rm TN})\) has vanishing mean when
\(d_</d_>\to0\) \cite{Page1993,Sen1996}.
In this regime the network realizes the sharp island formula
\cite[eq.~(5.14)]{ASS2023},
\begin{equation}
 S(\rho_A^{\rm TN})
 =\min\!\left\{\sum_{i\in A}S_i,\sum_{i\notin A}S_i\right\}+o(1),
 \qquad S_i=\log b_i.
 \label{eq:network-island-formula}
\end{equation}
When the two dimensions are comparable, the Page correction remains
of order one and the minimum gives only the leading entropy.

The two sums correspond to the two bulk regions considered in
section~\ref{sec:island}.  Before contraction with the shell tensor,
the state consists of independent entangled pairs
\(|\Phi_i\rangle_{a_i c_i}\) and the pure probe state
\(|\chi\rangle_{c_0}\).
For the region \(a_A\), each pair with \(i\in A\) has one end inside
and one outside, giving entropy \(\sum_{i\in A}S_i\).
For the region \(a_A\) together with the entire center, both ends
of these pairs lie inside.  Only the pairs with \(i\notin A\)
cross between the region and its complement, giving
\(\sum_{i\notin A}S_i\).
The independent pure state on \(c_0\) contributes no entropy.
These are two choices of region in the same state, not operations
that change the state.

At leading order, the candidate containing the entire center has
lower entropy when \(B_A>B_{\bar A}\), so the center is the network
island for \(A\).
The minimum becomes a sharp approximation when
\(|\log B_A-\log B_{\bar A}|\gg1\).
The competition is determined by the entanglement of the page pairs,
with no additional area term in this model.

The rank of the map instead concerns the full input space.
Since \(L^\dagger L=I\), the same
matrix gives the Wishart Gram operator and rank discussed in section~\ref{sec:rank-resolvent} and Appendix~E of~\cite{ASS2023},
\begin{equation}
 V_O^\dagger V_O=\frac1B X^\dagger X,\qquad
 \rank V_O=\min\{K_\psi,B\}\quad\text{almost surely}.
 \label{eq:network-wishart-rank}
\end{equation}
The resolvent follows from eq.~\eqref{eq:microcanonical-resolvent}
with \(D_\partial=B\).
For \(K_\psi>B\), some combinations of inputs have zero image.
Nevertheless, a collection of states fixed before the tensor is drawn
can have almost unchanged pairwise overlaps.  The null directions depend
on the draw and need not lie among that collection.
With the covariance above, the mean and overlap variance in~\cite[eqs.~(5.22)--(5.23)]{ASS2023} become
\begin{align}
 \mathbb E_{\rm G}(V_O^\dagger V_O)&=I_{K_\psi},
 \label{eq:network-mean-isometry}
 \\
 \mathbb E_{\rm G}
 \left|\langle u|(V_O^\dagger V_O-I)|v\rangle\right|^2
 &=\frac{\|u\|_2^2\|v\|_2^2}{B},
 \qquad u,v\in\mathcal H_\psi.
 \label{eq:network-overlap-variance}
\end{align}
For any preselected pair of normalized inputs, the mean-square overlap
error is therefore suppressed by \(B^{-1}\).

For a family \(\mathcal S=\{|u_a\rangle\}_{a=1}^M\) of normalized states
chosen independently of the tensor, define
\(\Delta_{\mathcal S}(V_O)=\max_{a,b}|\langle u_a|
(V_O^\dagger V_O-I)|u_b\rangle|\).  Gaussian concentration, as used in~\cite{ASS2023}, gives the sufficient estimate
\begin{equation}
 \mathbb P_{\rm G}[\Delta_{\mathcal S}(V_O)\geq2\delta]
 \leq4M^2e^{-B\delta^2/6},\qquad 0<\delta<1,
 \label{eq:network-test-family-bound}
\end{equation}
with these constants derived in appendix~\ref{app:resolvent}.
The selected overlaps are therefore preserved with error tending to
zero in probability if
\(\log M(B)=o(B^{1-2\gamma})\), with \(0<\gamma<1/2\) and
\(\delta=B^{-\gamma}\).  This is the approximate-isometry statement for a preselected family in~\cite{ASS2023}.  The estimate controls those
pairwise overlaps, rather than every superposition in their span.
\subsection{Three-leg PETS and multipartite entanglement}
\label{sec:tripartite-pets-entanglement}

For the fixed three-page PETS of
eq.~\eqref{eq:sg-three-leg-eth-state}, we can also ask what entanglement
local operations can extract.  Known results on random pure states
answer this question in its flat Gaussian specialization.

In a product energy window of dimensions \(b_1,b_2,b_3\), take the
ETH envelope and thermal weights to be constant and the coefficients
\(R_{abc}\) to be iid circular Gaussian variables.
The normalized state is then
\begin{equation}
 |\Psi_R\rangle_{123}
 =\frac{\sum_{a,b,c}R_{abc}|a,b,c\rangle}
 {\left(\sum_{a,b,c}|R_{abc}|^2\right)^{1/2}}.
 \label{eq:flat-three-leg-pure-state}
\end{equation}
The normalized coefficient vector is Haar distributed, and each sample
is a pure three-page state.\footnote{Nonconstant thermal weights
generally change the Haar distribution, even when the corresponding
filters are invertible.  The results quoted here use flat weights.}
Denote its one- and two-page reductions by $\rho_i^R$ and $\rho_{ij}^R$.
For three equal arms $b_i=b$, eq.~\eqref{eq:network-page-entropy} gives
an average single-page entropy deficit of order $b^{-1}$.  Markov's
inequality and global purity then imply, in probability as $b\to\infty$,
\begin{align}
 S(\rho_i^R)&=\log b+o(1),\qquad S(\rho_{ij}^R)=\log b+o(1),\nonumber\\
 I(i:j)&:=S(\rho_i^R)+S(\rho_j^R)-S(\rho_{ij}^R)=\log b+o(1).
 \label{eq:three-leg-flat-entropies}
\end{align}
Two pages therefore have substantial mutual information, with the same
leading entropies as the $b$-level Greenberger--Horne--Zeilinger (GHZ)
state $b^{-1/2}\sum_{j=1}^b|j,j,j\rangle$.
To characterize the entanglement beyond these entropies, consider what
the pages can extract by local operations.

An EPR pair is a maximally entangled pair of qubits.
Consider extracting $m\geq1$ such pairs from one copy of $\rho_{ij}^R$
by deterministic local channels on pages $i$ and $j$, without
communication or assistance from the traced-out third page.
Let \(F\) be the overlap of the output state with \(m\) EPR pairs
shared between two local \(m\)-qubit output systems, so $F$ is the squared fidelity.
Fix $m$ and a target $F\geq F_0$ with $1/2<F_0<1$.
Set \(b_i=2^{n_i}\), \(n=\sum_i n_i\), and require
\(n_i\leq(1/2-\eta)n\) for fixed \(\eta>0\).
Li, Mori and Yoshida show that, as $n\to\infty$, the probability that
any such local channels achieve the target is exponentially small in
the total Hilbert-space dimension \cite[theorem~2]{LiMoriYoshida2026}.
This includes local unitaries followed by tracing out the
remaining subsystems.
Their theorem~5 further excludes, with
probability \(1-o(1)\), exact extraction by local unitaries of a
nontrivial smaller pure tripartite factor, including a smaller GHZ
factor \cite{LiMoriYoshida2026}.

Substantial three-page correlations therefore coexist with a strong
restriction on local EPR extraction.  The local-unitary result also
distinguishes a typical sample from a GHZ state,
despite their matching leading entropies.

A fixed three-leg tensor can also be read as a map from leg~3 to
legs~1 and~2.  On its positive Gram support, polar normalization gives
an isometry, so access to~12 gives access to \emph{all} outputs and
admits an exact left inverse.
The booklet erasure problem instead uses the independent input \(c_0\)
of eq.~\eqref{eq:network-gaussian-map}, encoded into three outputs.
Losing access to output~3 then leaves only part of the encoded state.
Section~\ref{sec:erasure-qec} asks when those two remaining pages suffice
to recover an unknown input, including its quantum coherence.
% ===== END sections/tensor_network =====
%
% ===== BEGIN sections/erasure_qec =====
\section{Access to cosmological information}
\label{sec:erasure-qec}

Can two boundaries jointly read and manipulate a central probe
without assistance from the third?  Unlike joint access to all outputs
in the isometric model \(C\to AB\) of Mori and Yoshida
\cite{MoriYoshida2026}, access to \(N_1N_2\) in a booklet
encoding \(C\to N_1N_2N_3\) leaves a real output unavailable.
Here \(N_i\) is boundary \(i\), and \(C\) is a chosen finite probe
space.
In the flat Gaussian network, a small code can be approximately hidden from each
single page and recovered approximately from every pair with high
probability.  We relate this result to the page entanglement and the
whole-cosmology island, extending the access question raised in~\cite{ASS2023}.

\subsection{Single-page privacy and two-page recovery}
\label{sec:qec-contract}
\label{sec:exact-qec}
\label{sec:approx-qec}
\label{sec:gaussian-qec}

Recovery must preserve an unknown probe state, including its phases
and entanglement with an untouched reference.  At a fixed finite
regulator, choose a \(K\)-dimensional code \(\Hc\) with isometric
inclusion \(J_C:\Hc\to\mathcal H_\psi\).  Restrict the preparation map
of section~\ref{sec:boundary-map} to this code and assume its Gram
operator is positive definite.  Polar normalization gives
\cite[definition~3.4]{MaTanWright2025}
\begin{equation}
 \VC=V_\psi J_C,\qquad \GC=\VC^\dagger\VC,\qquad
 \WC=\VC\GC^{-1/2},\qquad \WC^\dagger\WC=I_C.
 \label{eq:paper-qec-restricted-map}
\end{equation}
\equationaliases{\label{eq:paper-qec-polar-isometry}}
The full boundary retains the input to \(W_C\) exactly.
This does not make \(V_C\) isometric.  Applied to a normalized state
prepared by \(V_C\), a decoder for \(W_C\) returns polar-normalized
probe coordinates.  They agree with the original coordinates
throughout the code only when \(G_C\) is proportional to the identity.
Appendix~\ref{app:approx-recovery} gives the coordinate relation.

For accessible pages \(A\) and inaccessible pages \(\bar A\), define
the retained and complementary channels
\cite{KretschmannKribsSpekkens2008}
\begin{equation}
 \cN_A(\rho)=\Tr_{\bar A}(W_C\rho W_C^\dagger),\qquad
 \cN_A^c(\rho)=\Tr_A(W_C\rho W_C^\dagger),
 \label{eq:paper-qec-erasure-channels}
\end{equation}
where \(\rho\) is a density operator on \(\Hc\).
Erasure removes access to a page, including its measurement records
or assistance, not the page itself from the geometry.
We use the half-diamond recovery error
\cite[definition~1.2]{MaTanWright2025},
\begin{equation}
 \epsilon_{\rm rec}(A)=\frac12\inf_{\cR_A\ {\rm CPTP}}
 \|\cR_A\circ\cN_A-\id_C\|_\diamond .
 \label{eq:paper-qec-recovery-error}
\end{equation}
The infimum is over completely positive, trace-preserving decoders
from the retained pages to \(C\).  The diamond norm tests the worst
input, including its reference entanglement
\cite[chapter~3]{Watrous2018}.
A decoder may depend on the known encoding but not on the unknown
input.  It may act jointly on the retained quantum systems. Separate
measurements and classical communication need not suffice.

Recovery is governed by the information left on the inaccessible
pages.  Exact correction requires their channel to be independent
of every logical input, including coherences
\cite{KnillLaflamme1997,KretschmannKribsSpekkens2008}.
Small departures from this condition permit approximate recovery.
The quantitative relation and norm conventions are collected in
appendix~\ref{app:approx-recovery}.

For the flat network of section~\ref{sec:tensor-network}, the fixed
restriction \(X_C=XJ_C\) remains an iid complex Gaussian matrix.
Its polar factor \(W_O=X_C(X_C^\dagger X_C)^{-1/2}\) is a
Haar-distributed orthonormal \(K\)-frame when \(K\leq B\)
\cite[lemma~3.6]{MaTanWright2025}.
The full boundary encoder is \(LW_O\), where the exterior isometries
\(L\) leave recovery unchanged.  We use a superscript \(O\) for
this model's channels and errors.  Thus the same junction tensor gives
the orthogonalized Gaussian random code of ref.~\cite{HSW2008};
no separate coding ensemble is introduced.  The Haar law uses flat
pair weights and is not asserted for general thermal weights.

Consider three equal arms of effective dimension \(b>1\), with
\(B=b^3\).  Each pair \(A_i=\{1,2,3\}\setminus\{i\}\) has dimension
\(b^2\), and the erased page has dimension \(b\).
For a code fixed before drawing the tensor, the results of Ma, Tan
and Wright \cite[theorems~1.10 and~3.1, lemma~3.5]{MaTanWright2025}
give
\begin{equation}
 \begin{split}
 &\zeta_b:=\frac{[\log(b^3)]^3}{b},\qquad
 r_b:=3\left[\sqrt{\frac Kb}
                 +C_*\bigl(\zeta_b^{1/4}+\zeta_b^{1/2}\bigr)\right],\\
 &\Prob_X\!\left[\max_{i=1,2,3}
 \epsilon_{\rm rec}^{O}(A_i)>r_b\right]
 \leq\min\!\left\{1,6\exp[-(\log b^3)^3]\right\},
 \qquad r_b<1.
 \end{split}
 \label{eq:paper-qec-three-page-rate}
\end{equation}
Here \(C_*>0\) is a universal constant.
The bound concerns all three pairs of the same encoding, each with
its own decoder.  Its probability is over encodings, not decoding
trials; no independence between pairs is assumed.
The ratio \(K/b\) compares the input dimension with the advantage
\(b^2/b\) of the retained pages.  Both fluctuation terms vanish as
\(b\to\infty\), so \(K/b\to0\) suffices for the error and the
probability of exceeding it to vanish.  The unspecified constant
precludes a numerical tolerance at given dimensions.
Appendix~\ref{app:gaussian-leakage} gives the source specialization
and the estimate for general page sets.

For each page \(i\), let \(\mathcal N_i^{O}\) denote the channel
that maps a logical input \(\rho\) to the reduced state of that
page after encoding.  Its complementary channel has access to
the other two pages.  If those two pages recover the input with
error at most \(r_b\), privacy--correctability complementarity
implies that there is a density operator \(\sigma_i\) such that
\cite[theorem~5]{KretschmannKribsSpekkens2008}
\[
 \frac12
 \bigl\|\mathcal N_i^{O}-\mathcal C_{\sigma_i}\bigr\|_\diamond
 \leq \sqrt{2r_b},
 \qquad
 \mathcal C_{\sigma_i}(\rho):=\sigma_i\operatorname{Tr}\rho .
\]
The replacement channel \(\mathcal C_{\sigma_i}\) returns the
same state for every normalized input.  The state \(\sigma_i\)
may depend on the known encoding, but not on the unknown probe
state.  Thus the actual output on page \(i\) is nearly independent
of the logical input.  The diamond-norm statement also controls
its correlations with an untouched reference.

Consider any CPTP decoder \(\mathcal R_i\) acting on this page.
Applying it to both channels cannot increase their diamond
distance.  Since
\(\mathcal R_i\circ\mathcal C_{\sigma_i}
=\mathcal C_{\tau_i}\), with
\(\tau_i=\mathcal R_i(\sigma_i)\), the decoded channel remains
within \(\sqrt{2r_b}\), in half-diamond distance, of a fixed-output
channel.  Such a channel cannot recover the input's entanglement
with a reference.

To quantify this obstruction, take the maximally entangled input
\[
 |\Phi_K\rangle
 =\frac1{\sqrt K}\sum_{I=1}^{K}|I\rangle_R|I\rangle_C .
\]
Ideal recovery preserves this state, whereas a fixed-output
channel gives \(I_R/K\otimes\tau_i\).  Measuring the projector
\(|\Phi_K\rangle\langle\Phi_K|\) therefore gives probabilities
\(1\) and
\[
 \langle\Phi_K|
 \left(\frac{I_R}{K}\otimes\tau_i\right)
 |\Phi_K\rangle
 =\frac{1}{K^2},
\]
respectively.  Their trace distance is consequently at least
\(1-K^{-2}\).  Any decoded single-page channel remains within \(\sqrt{2r_b}\)
of a fixed-output channel, whose half-diamond distance from the
identity is at least \(1-K^{-2}\).
The triangle inequality therefore bounds its distance from the
identity below by \(1-K^{-2}-\sqrt{2r_b}\), for every decoder.
\begin{equation}
 \epsilon_{\rm rec}^{O}(\{i\})
 \geq\max\{0,1-K^{-2}-\sqrt{2r_b}\},
 \qquad i=1,2,3 .
 \label{eq:paper-privacy-recovery-lower}
\end{equation}

This conclusion holds on the same high-probability event as the
three pair-recovery bounds in
eq.~\eqref{eq:paper-qec-three-page-rate}.  For a logical qubit,
the single-page recovery error is at least \(3/4-o(1)\) as
\(b\to\infty\) with \(K/b\to0\), while every pair's error tends
to zero.  The same encoding can therefore hide the probe from
each individual page while allowing its recovery by joint
quantum operations on any pair.  This is an approximate
quantum-secret-sharing pattern \cite{CGL1999}.
We next examine how this access depends on the entanglement
assigned to the pages.
\subsection{Entanglement, code size and the whole-center island}
\label{sec:access-island-entanglement}

In the large-mass pair approximation, exterior \(a_i\) is entangled
with cosmological patch \(c_i\), with entropy
\(S_i=S(\rho_{a_i})=S(\rho_{c_i})\).
A region containing only \(a_A\) cuts every accessible page pair.
Including the whole cosmology puts both ends of those pairs inside
the region, leaving only the pairs attached to inaccessible exteriors
across its boundary.  The entropy comparison
\cite[section~3.3]{ASS2023} therefore becomes, at this order,
\begin{equation}
 S_{\rm no\ island}(A)=\sum_{i\in A}S_i,\qquad
 S_{\rm whole\ cosmology}(A)=\sum_{i\notin A}S_i.
 \label{eq:paper-access-island-entropies}
\end{equation}
Both candidates have an empty extremal surface, so there is no
additional area term in their competition.
Under the saddle assumptions of section~\ref{sec:island}, away
from the transition, the smaller entropy selects the region.
For three equal pair entropies \(s>0\), one page compares \(s\)
with \(2s\) and excludes the center; two pages compare \(2s\)
with \(s\) and include it.  The center enters as a whole, without
assigning each \(c_i\) exclusively to its corresponding boundary.

The flat reference pairs have \(S_i=\log b_i\).
Here \(b_i\) is an effective entangled-arm dimension, not the full
CFT dimension or a geometric volume.  Put
\(B_A=\prod_{i\in A}b_i\), \(B_{\bar A}=\prod_{i\notin A}b_i\),
and \(B=B_AB_{\bar A}\).  The reference entropy advantage and
the code-dependent ratio in the Haar recovery bound are related by
\begin{align}
 &\Delta_A:=\sum_{i\in A}S_i-\sum_{i\notin A}S_i
           =\log\frac{B_A}{B_{\bar A}},
 \label{eq:paper-access-entanglement-balance}\\
 &\frac{KB_{\bar A}}{B_A}=\exp(\log K-\Delta_A).
 \label{eq:paper-access-code-ratio}
\end{align}
These are reference-pair entropies, not the marginal entropies of
a Gaussian boundary realization.
The whole-center candidate is favored by \(\Delta_A>0\), whereas
uniform recovery must accommodate an entire \(K\)-dimensional input
and its reference entanglement.
The general estimate in appendix~\ref{app:gaussian-leakage}
\cite{MaTanWright2025} gives the sufficient regime
\begin{equation}
 \frac{KB_{\bar A}}{B_A}\longrightarrow0,\qquad
 \frac{B_{\bar A}(\log B)^3}{B_A}\longrightarrow0
 \label{eq:paper-qec-gaussian-sufficient-regime}
\end{equation}
\equationaliases{\label{eq:intro-gaussian-scaling}}
for recovery error tending to zero in probability.
The second condition controls finite-dimensional fluctuations.
The statement also holds for a fixed finite family of page sets
chosen before the draw. Neither condition is a necessary threshold.

For three equal arms, this reduces to \(K/b\to0\) as \(b\to\infty\).
Enlarging the low-energy probe space asks a decoder to handle more
states and superpositions, without necessarily changing the background.
A two-page island therefore does not grant uniform access to an
arbitrarily large code, as is familiar from code-space-dependent
holographic reconstruction \cite{HaydenPenington2019}.
Large shell mass, the semiclassical limit and large arm dimensions
are separate assumptions in this comparison.

\subsection{Approximate recovery and its holographic interpretation}
\label{sec:logical-access}
\label{sec:qec-exactness-physical}
\label{sec:physical-qec}

The recovery results give a finite-dimensional model of how central
quantum information can remain accessible when one boundary is
unavailable.  In the flat Gaussian network, the bounds of the
preceding sections guarantee accurate recovery on a preselected
probe space in the stated large-dimension regime.  Their
holographic significance lies in the possibility of an approximate
boundary description of these probes, rather than an exact
implementation of all central operations.

The recovery error has a direct meaning for observables.
Suppose a decoder satisfies
\(\frac12\|\mathcal R_A\circ\mathcal N_A^{O}
-\operatorname{id}_C\|_\diamond\leq\varepsilon\).
For a Hermitian observable \(O_C\) on the chosen code, with
\(\|O_C\|_\infty\leq1\), the adjoint recovery map defines an
observable on the retained pages
\cite{CotlerEtAl2019},
\begin{equation}
 \begin{split}
 &O_A:=\mathcal R_A^\dagger(O_C),\\
 &\left|
 \operatorname{Tr}\!\left[O_A\mathcal N_A^{O}(\rho)\right]
 -\operatorname{Tr}(O_C\rho)
 \right|\leq2\varepsilon .
 \end{split}
 \label{eq:paper-approximate-probe-observable}
\end{equation}
The adjoint is defined by
\(\operatorname{Tr}[O_C\mathcal R_A(\omega)]
=\operatorname{Tr}[\mathcal R_A^\dagger(O_C)\omega]\).
The bound follows from trace-norm duality and holds for every
density operator \(\rho\) on the code, with the same \(O_A\).
Thus the retained pages can reproduce central expectation values
to controlled accuracy.  The diamond-norm condition also tests
preservation of correlations with an untouched reference.
This does not require an exact identification of the bulk and
boundary operator algebras.  Conditions for implementing a
specified logical operation are discussed separately in
appendix~\ref{app:approx-recovery}.

This interpretation is compatible with the whole-center island
of section~\ref{sec:access-island-entanglement}.
The island calculation compares the leading entropies of two
candidate regions for a reference state.  Recovery instead asks
whether one decoder works throughout a specified probe space.
A preference for including the whole center can therefore coexist
with small distinctions between probe states on the inaccessible
page. Residual probe information on the inaccessible page obstructs
exact recovery of the entire code from the other two pages.
This can coexist with a leading entropy comparison that favors
the whole-center island, provided the corrections do not reverse
the ordering of the two candidate entropies. The full encoded state remains pure for a pure
input.  Imperfect recovery from two pages concerns restricted
access to its correlations, not a loss of information from the
full boundary or a failure of unitary evolution.

The distinction between approximate and exact recovery also
separates the Gaussian model from perfect-tensor constructions.
HaPPY achieves exact reconstruction in its correctable regions
through special tensor identities
\cite[sections~5.3--5.4]{PastawskiEtAl2015}.
The Gaussian model instead uses high-dimensional typicality to
suppress the input dependence of the erased subsystem.
For a fixed erasure and a preselected nontrivial code, exact
Knill--Laflamme conditions hold only on a zero-measure set of
encoders \cite{KnillLaflamme1997,MaTanWright2025},
\begin{equation}
 \Prob_X[\epsilon_{\rm rec}^{O}(A)=0]=0,
 \qquad
 2\leq K\leq B_AB_{\bar A},\quad B_{\bar A}\geq2 .
 \label{eq:paper-qec-generic-no-exact}
\end{equation}
\footnote{
Let \(w_1,w_2\) be two columns of the Haar encoder and
\(P=I_{B_A}\otimes|1\rangle\langle1|\).
Exact correction requires
\(w_1^\dagger Pw_1=w_2^\dagger Pw_2\).
Their difference is a real analytic function on the connected
manifold of orthonormal \(K\)-frames.  It is not identically zero,
as the columns \(|1\rangle_A|1\rangle_{\bar A}\) and
\(|1\rangle_A|2\rangle_{\bar A}\) give difference one.
Its zero set has Haar measure zero.
The infimum over finite-dimensional CPTP decoders is attained,
so zero recovery error requires exact correction.
}
This finite-dimensional statement does not impose a nonzero
asymptotic error floor.  Exact correction can have probability
zero at every finite dimension while the recovery error tends
to zero in the regime established above.

To establish the same recovery property for the physical booklet,
one must determine how much information about the central probe
remains on the inaccessible boundaries.
For the polar encoder obtained from the same Euclidean preparation,
the required condition is that these boundaries become insensitive
to the logical input.
More precisely, their channels must approach input-independent
channels in diamond norm, as specified in
eq.~\eqref{eq:app-physical-recovery-premise}.
This controls all states in the chosen code, also when the probe
is entangled with a reference, and must hold uniformly over the
boundary subsets considered.
Privacy--correctability complementarity then implies approximate
recovery from the retained boundaries
\cite{KretschmannKribsSpekkens2008}.

This conclusion concerns the polar encoder \(W_C\), rather than
the original preparation map \(V_C\).
The relation between their input states is given in
appendix~\ref{app:approx-recovery}.
Neither the reference-state island calculation nor the Gaussian
model establishes the required decoupling for a fixed CFT.
The present results therefore provide a controlled model of
approximate access to central information and explain its
compatibility with a whole-center island.
Whether the physical Euclidean preparation satisfies the same
recovery condition remains to be determined.
% ===== END sections/erasure_qec =====
%
% ===== BEGIN sections/discussion =====
\section{Discussion}
\label{sec:discussion}

We extend the cosmological-state construction in~\cite{ASS2023} to the booklet geometry, where
multiway junction conditions determine the shell geometry and its Lorentzian continuatio~\cite{ShenPengLi2024}. 
In our construction, a crucial ingrediant is the three-leg tensor which helps separate entanglement from access to quantum
information.  Its flat Gaussian state is genuinely multipartite: under the
restrictions of ref.~\cite{LiMoriYoshida2026}, its correlations cannot be
reduced to independent EPR pairs. 
On the other hand, by another theorem~\cite[Theorems~5]{LiMoriYoshida2026} no exact GHZ factor can be extracted by local unitaries.  Therefore it is interesting to understand better the microscopic nature of this  state. In particular, apart from erasure, de-phasing effects~\cite{Mori:2026kgj} could also be a useful diagnose. Another interesting question is to extend our Haar discussion to include thermal weights. This requires additional input and we hope to present more results in this direction in the near future.

\paragraph{The use of AI tools}

LLM-based tools were used in preparing and polishing portions of the draft and in assisting with some technical computations. The scientific ideas, conceptual framework, and overall approaches pursued in this work were developed by the authors, and all AI-assisted computations relevant to the results presented here were independently verified by the authors. More specifically, in the fall of 2025, inspired by~\cite{Antonini:2025ioh}, CP developed the idea of extending the framework of~\cite{Antonini:2025ioh} to multiway-junction constructions and booklet geometries, together with possible applications to new holographic analogues of quantum information processes, including quantum error correction. In December 2025, CP communicated this direction, together with an initial roadmap for pursuing it, separately to Prof. Yan Liu and JD in discussions concerning possible collaboration. JD and BD subsequently began developing the calculations presented in this manuscript. By June 2026, CP became aware that Prof. Yan Liu’s group was pursuing along this direction independently. JD and BD continued the development of the present work and carried out the main body of the technical calculations. In the process, LLM-based tools are used for some detailed computations and JD and BD have explicitly verified the resulting calculations. CP independently cross checked these results and was primarily responsible for organizing the presentation and finalizing the manuscript. Noticing the work arXiv:2609.37099 posted and in view of the overlap in the multi-way junction construction, although the technical details and the results are significantly different, we have chosen to make the present version available at this stage, while leaving several calculations that remain in progress to a subsequent revision or a future publication.

\acknowledgments

The work of CP is supported by NSFC NO. 12675092, 12175237,  12447108, and in part by NO. 12247103, the Fundamental Research Funds for the Central Universities, and funds from the Chinese Academy of Sciences. The work of JD is supported in part by the NSFC under Grant No. 12475090. The work of BD is supported in part by NSFC under Grant No. 12275275, 12675082.
%
% ===== BEGIN appendices/saddle_competition =====

\appendix
\section{Normalization and fluctuations of the booklet map}
\label{app:boundary-network-support}
\label{app:resolvent}

\subsection{Overlap normalization}

The normalized overlaps in section~\ref{sec:gram-moments} use the
same norm saddles as in~\cite[eqs.~(4.12)--(4.19)]{ASS2023}.
For a type-I probe with particle masses
\(\mu_{I,\alpha}\simeq\Delta_{I,\alpha}/\ell\), its worldline
factor \(p_I\) satisfies
\begin{equation}
 \frac{p_I[X_n]}{p_I[X]}
 \simeq\exp\!\left[-\sum_\alpha\mu_{I,\alpha}
 \bigl(L^{\rm ren}_{I,\alpha}[X_n]
       -L^{\rm ren}_{I,\alpha}[X]\bigr)\right].
 \label{eq:gram-probe-length-ratio}
\end{equation}
Here \(X\) is the single-copy saddle and \(X_n\) the connected
cyclic saddle for the overlap moment.  The selected endpoints have
separation \(2\tau_*\) in both geometries, where \(\tau_*\) is their
Euclidean half-separation.  Their common renormalized length is
\(2\ell\log[2\sinh(\tau_*/\ell)]\), so the leading worldline factors
cancel.  Together with the assumed cancellation of probe fluctuation
factors and unchanged reference page weights, this gives
\begin{equation}
 [|G_{IJ}|^2]_{\rm sc,conn}
 \simeq\frac{p_I[X_2]p_J[X_2]}{p_I[X]p_J[X]}
       \prod_iq_i(2)
 \simeq\prod_iq_i(2),\qquad I\ne J .
 \label{eq:gram-probe-normalization-cancellation}
\end{equation}
The same assumptions apply at each fixed higher moment order.
The probe dimensions and combined backreaction obey
section~\ref{sec:gram-moments}.

The Gaussian calculation distinguishes raw from individually
normalized images.  Put
\(\Omega=\bigotimes_i\rho_i^{\rm ref}\), with \(\Tr\Omega=1\), and let
\(|r_I\rangle\) be independent standard complex Gaussian vectors.
For \(\widehat G^{\rm G}_{IJ}=\langle r_I|\Omega|r_J\rangle\)
and pairwise distinct \(I_1,\ldots,I_n\), the connected Wick contraction
follows the whole cycle,
\begin{equation}
 \mathbb E\!\left[\widehat G^{\rm G}_{I_1I_2}\cdots
 \widehat G^{\rm G}_{I_nI_1}\right]_{\rm conn}
 =\Tr\Omega^n=\prod_i\Tr[(\rho_i^{\rm ref})^n].
 \label{eq:app-connected-correlated-moment}
\end{equation}
All proper-subset moments vanish.  If instead
\(|v_I\rangle=\Omega^{1/2}|r_I\rangle/
\sqrt{\langle r_I|\Omega|r_I\rangle}\) and
\(M=\mathbb E[|v_I\rangle\langle v_I|]\), independence gives
\begin{equation}
 \mathbb E\!\left[
 \langle v_{I_1}|v_{I_2}\rangle\cdots
 \langle v_{I_n}|v_{I_1}\rangle\right]=\Tr M^n.
 \label{eq:app-normalized-gaussian-cycle}
\end{equation}
The product is the trace of \(n\) independent projectors.
In general \(M\ne\Omega\).  For a flat window
\(\Omega=P/D_\partial\), with \(P\) a rank-\(D_\partial\) projector,
unitary invariance gives \(M=\Omega\) and both cycles equal
\(D_\partial^{1-n}\).  Thus the Gaussian cycle structure alone does
not replace the semiclassical norm prescription for general weights.

The independent entries in section~\ref{sec:common-network} also use
the stated choice of central inputs.  In the more general fixed contraction of eq.~(5.17) in~\cite{ASS2023}, write the
projected input as
\(\sum_{\boldsymbol\alpha,\mu}\phi^I_{\boldsymbol\alpha\mu}
|\boldsymbol\alpha\rangle_a|\mu\rangle_{\rm cen}\), and the shell
bra as \(\sum_\mu O_\mu\langle\mu|\).  The induced entries are
\(X^{\rm gen}_{\boldsymbol\alpha I}
=\sum_\mu O_\mu\phi^I_{\boldsymbol\alpha\mu}\).  For shell covariance
\(\Gamma_{\mu\nu}=\mathbb E[O_\mu O_\nu^*]\),
\begin{equation}
 \mathbb E[X^{\rm gen}_{\boldsymbol\alpha I}
 X^{\rm gen*}_{\boldsymbol\beta J}]
 =\sum_{\mu,\nu}\phi^I_{\boldsymbol\alpha\mu}
 \Gamma_{\mu\nu}\phi^{J*}_{\boldsymbol\beta\nu}.
 \label{eq:network-general-induced-covariance}
\end{equation}
The independent input factor \(c_0\), flat pairs and Gaussian shell
of section~\ref{sec:common-network} give its iid specialization.
Equal exterior marginals alone do not imply independent columns.

\subsection{Fluctuations for a preselected family}

The rank and planar resolvent are given in
section~\ref{sec:rank-resolvent}.  Here we derive the finite-family
constant in eq.~\eqref{eq:network-test-family-bound}.
Put \(A_O=V_O^\dagger V_O-I\) and
\(Q_O(u)=\langle u|A_O|u\rangle\).  For a fixed unit vector,
\(Q_O(u)=B^{-1}\sum_{\alpha=1}^B(Y_\alpha-1)\), where the
\(Y_\alpha\) are independent unit-mean exponential variables.
Their generating function
\(\mathbb E e^{t(Y-1)}=e^{-t}/(1-t)\) gives Chernoff exponents
\(\delta-\log(1+\delta)\) and
\(-\delta-\log(1-\delta)\).  Both are at least \(\delta^2/6\)
for \(0<\delta<1\), hence
\begin{equation}
 \mathbb P[|Q_O(u)|\geq\delta]\leq2e^{-B\delta^2/6}.
 \label{eq:appendix-fixed-vector-bound}
\end{equation}
Complex polarization gives
\begin{equation}
 4\langle u|A_O|v\rangle
 =\sum_{r=0}^3(-i)^r
   \langle u+i^rv|A_O|u+i^rv\rangle .
 \label{eq:appendix-complex-polarization}
\end{equation}
For the \(M\) unit vectors in the preselected family, apply the
diagonal bound also to every nonzero normalized
\(u_a+i^ru_b\), \(a<b\).  There are at most \(2M^2\) test vectors.
Since \(\sum_r\|u+i^rv\|^2=8\), the resulting overlap error
is at most \(2\delta\).  A union bound yields
\begin{equation}
 \mathbb P[\Delta_{\mathcal S}(V_O)\geq2\delta]
 \leq4M^2e^{-B\delta^2/6},
 \label{eq:appendix-finite-set-bound}
\end{equation}
for the same Gaussian draw and a family chosen independently of it.

\section{Recovery criteria and logical operations}
\label{app:approx-recovery}

\paragraph{Preparation and polar coordinates.}
For a normalized probe \(|\chi\rangle\), polar decomposition gives
\begin{equation}
 \frac{V_C|\chi\rangle}{\|V_C|\chi\rangle\|}
 =W_C|\chi_{\rm pol}\rangle,\qquad
 |\chi_{\rm pol}\rangle
 =\frac{G_C^{1/2}|\chi\rangle}
 {\sqrt{\langle\chi|G_C|\chi\rangle}} .
 \label{eq:paper-qec-prepared-state-coordinates}
\end{equation}
A decoder for \(W_C\) returns \(|\chi_{\rm pol}\rangle\).
These coordinates agree with the original ones over the whole code
when \(G_C\) is proportional to the identity.  Approximate agreement
requires uniform Gram control.  The same logical unitary acts in
both coordinates when \([G_C,U_C]=0\).

\paragraph{Leakage and recovery.}
We use \(\|X\|_1=\Tr\sqrt{X^\dagger X}\).
For a difference of channels with input \(C\), the diamond norm
maximizes the output trace norm over density operators on
\(C\otimes R\), where \(\dim R=K\)
\cite[theorem~3.51]{Watrous2018}.
The complementary leakage is
\begin{equation}
 \eta(A):=\frac12\inf_{\sigma_{\bar A}}
 \|\cN_A^c-\mathcal C_{\sigma_{\bar A}}\|_\diamond,\qquad
 \mathcal C_\sigma(\rho)=\sigma\Tr\rho ,
 \label{eq:paper-qec-leakage}
\end{equation}
where the infimum is over density operators on the erased system.
Exact correction is equivalent to \(\eta(A)=0\)
\cite{KnillLaflamme1997,KretschmannKribsSpekkens2008}.
In the half-diamond convention, theorem~5 of
ref.~\cite{KretschmannKribsSpekkens2008} gives
\begin{equation}
 \frac{\eta(A)^2}{2}\leq\epsilon_{\rm rec}(A)
 \leq\min\{1,\sqrt{2\eta(A)}\}.
 \label{eq:paper-qec-information-disturbance}
\end{equation}
Thus the physical application in section~\ref{sec:physical-qec}
requires a family of the same-preparation maps \(V_\psi^{(\nu)}\),
preselected code inclusions \(J_C^{(\nu)}\), and positive
\(G_C^{(\nu)}\), for which
\begin{equation}
 \sup_{A\in\mathcal F}\eta_{\rm phys}^{(\nu)}(A)\longrightarrow0
 \quad\Longrightarrow\quad
 \sup_{A\in\mathcal F}\epsilon_{\rm rec,phys}^{(\nu)}(A)
 \longrightarrow0 .
 \label{eq:app-physical-recovery-premise}
\end{equation}
Each \(\nu\) denotes a finite-regulator preparation and
\(\mathcal F\) is a fixed finite family of accessible page sets.
The implication is a channel statement, conditional on the leakage
premise rather than on the reference-state island alone.

For the two inputs in eq.~\eqref{eq:discussion-leakage-witness},
the triangle inequality gives
\(D_{\mathrm{tr}}^{(3)}\leq2\eta(\{1,2\})\).
Equation~\eqref{eq:paper-qec-information-disturbance} then gives
\(\epsilon_{\rm rec}(\{1,2\})\geq
(D_{\mathrm{tr}}^{(3)})^2/8\).
A Hermitian observable with \(\|O_3\|_\infty\leq1\) obeys
\(|\Tr[O_3(\rho_3^{(0)}-\rho_3^{(1)})]|
\leq2D_{\mathrm{tr}}^{(3)}\), as used in the discussion.

\paragraph{Operations on two pages.}
Let \(U_C\) be a unitary preserving the chosen code, and define
\(\mathcal N_3:=\mathcal N_{\{1,2\}}^c\).
An exact implementation on the first two pages would require a
unitary \(\widetilde U_{12}\) satisfying
\begin{equation}
 (\widetilde U_{12}\otimes I_3)W_C=W_CU_C .
 \label{eq:paper-two-page-unitary}
\end{equation}
This is a condition on the specified encoder and operation,
not a property assumed for the physical booklet.
Since a unitary acting on the first two pages cannot change
the third-page reduced state, it requires
\begin{equation}
 \mathcal N_3(U_C\rho U_C^\dagger)=\mathcal N_3(\rho)
 \quad\text{for every density operator \(\rho\) on \(\mathcal H_C\)}.
 \label{eq:paper-gate-complement-condition}
\end{equation}
Under eq.~\eqref{eq:paper-gate-complement-condition},
\(W_C\) and \(W_CU_C\) dilate the same channel to \(N_3\).
Stinespring uniqueness identifies their minimal dilation supports
by an isometry \cite[section~III.2]{KretschmannEtAl2008}.
The supports have equal dimension inside \(\mathcal H_{12}\),
so that isometry extends to a unitary on \(\mathcal H_{12}\).
This proves eq.~\eqref{eq:paper-two-page-unitary};
tracing over \(\mathcal H_{12}\) proves the converse.

For the approximate case, define
\(\mathcal U_C(\rho)=U_C\rho U_C^\dagger\) and
\begin{equation}
 \nu_U:=\frac12
 \|\mathcal N_3\circ\mathcal U_C-\mathcal N_3\|_\diamond
 \leq2\eta(\{1,2\}).
 \label{eq:app-gate-leakage}
\end{equation}
The inequality follows by inserting a replacement channel, which is
unchanged by \(\mathcal U_C\), and using the triangle inequality.
Append an accessible auxiliary qubit \(F\) in state \(|0\rangle_F\)
to obtain \(W_0:C\to\mathcal H_{12}\otimes\mathcal H_F\otimes\mathcal H_3\).
The doubled accessible space is large enough for the Stinespring
continuity theorem \cite{KretschmannEtAl2008,vomEnde2023}.
It gives a unitary \(\widetilde U_{12F}\) with
\begin{equation}
 \|(\widetilde U_{12F}\otimes I_3)W_0-W_0U_C\|_\infty
 \leq\sqrt{2\nu_U},
 \label{eq:app-gate-dilation-bound}
\end{equation}
where \(\|\cdot\|_\infty\) is the operator norm.
Writing \(\mathcal W_0(\rho)=W_0\rho W_0^\dagger\) and
\(\widetilde{\mathcal U}_{12F}(X)=
\widetilde U_{12F}X\widetilde U_{12F}^\dagger\), define
\begin{equation}
 \begin{split}
 \epsilon_{\rm op}(U_C)
 &:=\frac12\inf_{\widetilde U_{12F}}
 \|(\widetilde{\mathcal U}_{12F}\otimes\id_3)\circ\mathcal W_0
       -\mathcal W_0\circ\mathcal U_C\|_\diamond\\
 &\leq\min\{1,\sqrt{2\nu_U}\}
 \leq\min\{1,2\sqrt{\eta(\{1,2\})}\}.
 \end{split}
 \label{eq:paper-gate-error}
\end{equation}
Expanding the difference of the two output density operators gives
the first bound from eq.~\eqref{eq:app-gate-dilation-bound}, also with
an arbitrary reference.  Both channels include \(F\), so the
comparison tests its return to \(|0\rangle_F\).
The operation may depend on the known encoder and \(U_C\), but not
on the unknown input.
\section{Gaussian erasure estimates}
\label{app:gaussian-leakage}

The restriction \(X_C=XJ_C\) to a code fixed before the draw remains
iid by Gaussian unitary invariance.  For \(K\leq B\), its polar
factor is a Haar \(K\)-frame
\cite[lemma~3.6]{MaTanWright2025}.
The exterior isometries \(L=\bigotimes_iU_i^{\rm ext}\) leave the
optimal erasure error unchanged.  A decoder on their retained
image extends to a CPTP map by sending its orthogonal complement
to a fixed state.

Put \(a=B_A\), \(e=B_{\bar A}\), and \(B=ae>1\).
To apply the unitary-error analysis of ref.~\cite{MaTanWright2025},
take the \(m=e^2\) trace-orthogonal Weyl unitaries
\(E_{rs}=I_a\otimes X_e^rZ_e^s\), \(0\leq r,s<e\), where
\(X_e|j\rangle=|j+1\bmod e\rangle\) and
\(Z_e|j\rangle=e^{2\pi i j/e}|j\rangle\).
Their average channel is
\begin{equation}
 \frac1{e^2}\sum_{r,s}E_{rs}\rho E_{rs}^\dagger
 =\Tr_{\bar A}\rho\otimes\frac{I_e}{e}.
 \label{eq:app-qec-weyl-erasure}
\end{equation}
It is erasure followed by appending a fixed state, so recovery
from this channel requires access only to \(A\).

Let \(T_X=\sum_{\mu=1}^{m}E_\mu(X_C/\sqrt B)\otimes\langle\mu|\)
map logical and error labels to their output vectors.
The variance matrices in theorem~3.1 and the proof of lemma~3.3
of ref.~\cite{MaTanWright2025} are
\(\mathbb E(T_X^\dagger T_X)=I_{Km}\) and
\(\mathbb E(T_XT_X^\dagger)=(Km/B)I_B\).
For \(Km\leq B\), their larger operator norm gives
\(\sigma(T_X)=1\), and the entry covariance gives
\(v(T_X)=\sqrt{m/B}\).
The source concentration theorem, with \(t=(\log B)^{3/2}\),
places all singular values of \(T_X\) between \(1-u\) and \(1+u\),
except on an event of probability at most \(2e^{-(\log B)^3}\), where
\begin{equation}
 \begin{split}
 \zeta_A&:=\frac{e(\log B)^3}{a},\\
 u_{a,e,K}&:=\sqrt{\frac{Ke}{a}}
        +C_*(\zeta_A^{1/4}+\zeta_A^{1/2}),\qquad r_A:=3u_{a,e,K}.
 \end{split}
 \label{eq:app-qec-haar-remainder}
\end{equation}
\equationaliases{\label{eq:paper-qec-haar-rate}}
Here \(u=u_{a,e,K}\), \(C_*>0\) is universal, and logarithms
are natural.  Both concentration terms are retained.\footnote{
In arXiv:2510.07158v1, the last simplification in the proof of
lemma~3.3 replaces \(x^{1/4}+x^{1/2}\) by \(2x^{1/2}\).
That inequality fails for \(0<x<1\).
We use the preceding unsimplified expression, not the rate printed
in theorem~1.11.}
If \(3u<1\), then \(Km<B\).
Lemma~3.5 bounds the singular-value error after polar normalization
by \(2u/(1-u)\leq3u\).  Theorem~1.10 then supplies one CPTP
decoder for all code states and their reference entanglement,
in the half-diamond convention of definition~1.2
\cite{MaTanWright2025},
\begin{equation}
 \Prob_X[\epsilon_{\rm rec}^{O}(A)>r_A]
 \leq\min\{1,2e^{-(\log B)^3}\},\qquad r_A<1 .
 \label{eq:app-qec-haar-bound}
\end{equation}

For a fixed finite family \(\mathcal F\) chosen before the draw,
with every \(r_A<1\), a union bound gives
\begin{equation}
 \Prob_X[\exists A\in\mathcal F:
              \epsilon_{\rm rec}^{O}(A)>r_A]
 \leq\min\{1,2|\mathcal F|e^{-(\log B)^3}\}.
 \label{eq:paper-qec-haar-probability}
\end{equation}
This gives eq.~\eqref{eq:paper-qec-three-page-rate} for the three pairs,
without assuming independent events.
For fixed \(\Npg\) equal arms and erasure of \(t\) pages, the two
ratios controlling \(u\) are
\(K/b^{\Npg-2t}\) and
\((\log b^{\Npg})^3/b^{\Npg-2t}\).
Thus \(t<\Npg/2\) and \(K/b^{\Npg-2t}\to0\), as \(b\to\infty\),
suffice.  Two equal arms with one erased certify no nontrivial code
by this estimate.  The cases \(K=1\) or \(e=1\) have exact recovery
without this bound.  Nonuniform thermal weights need not give Haar
encoders, so the estimate is restricted to the flat model.

\bibliographystyle{JHEP}
\bibliography{references}
\end{document}